\documentclass[10pt, letterpaper,twocolumn]{article}

\usepackage[margin=0.70in]{geometry}
\usepackage[affil-it]{authblk}
\usepackage[utf8]{inputenc}
\usepackage{filecontents}

\usepackage{xcolor}

\usepackage{lipsum}
\usepackage{graphicx}
\usepackage{float}
\graphicspath{{Figures/}} 
\usepackage{lmodern,amsmath,amsthm}

\usepackage{amsfonts}
\usepackage[bbgreekl]{mathbbol}
\DeclareSymbolFontAlphabet{\mathbbm}{bbold}
\DeclareSymbolFontAlphabet{\mathbb}{AMSb}

\usepackage{amssymb}
\usepackage[autopunct = true, citestyle=numeric-comp, autocite=superscript, style=nature, doi=false, isbn=false, url=false, maxbibnames=3]{biblatex}
 \newbibmacro{string+url}[1]{%
  \iffieldundef{url}{}{\href{\thefield{url}}{#1}}%
}

\DeclareFieldFormat{title}{\usebibmacro{string+url}{\mkbibemph{#1}}}
\DeclareFieldFormat[article,incollection]{title}%
    {\usebibmacro{string+url}{\mkbibquote{#1}}}
 
\usepackage[colorlinks]{hyperref}
\hypersetup{
     colorlinks   = true,
     allcolors    = blue
}
\usepackage[font=small,labelfont=bf]{caption}
\usepackage{titlesec}
\usepackage{nccmath} 

\usepackage{xr}
\makeatletter
\newcommand*{\addFileDependency}[1]{
  \typeout{(#1)}
  \@addtofilelist{#1}
  \IfFileExists{#1}{}{\typeout{No file #1.}}
}
\makeatother

\newcommand*{\myexternaldocument}[1]{
    \externaldocument{#1}
    \addFileDependency{#1.tex}
    \addFileDependency{#1.aux}
}

\myexternaldocument{Supplemental}

\titlespacing*\section{6pt}{12pt}{5pt}
\titlespacing*\subsection{6pt}{0pt}{5pt}
\titleformat*{\section}{\large\bfseries}
\titleformat*{\subsection}{\normalsize\bfseries}

\newcommand{\V}{\mathbb{V}}

\newcommand{\T}{\mathbb{T}}
\newcommand{\G}{\mathbb{G}}
\newcommand{\A}{\mathbb{A}}

\newcommand{\rpos}{\mathbf{r}}

\newcommand{\khat}{\hat{\mathbf{k}}}
\newcommand{\kind}{\khat^{\mathrm{s}}}

\newcommand{\Esrc}{\mathbf{E}^{\mathrm{i}}}

\newcommand{\ksrc}{\khat^{\mathrm{i}}}

\newcommand{\Asym}[1]{\mathrm{Asym}\left(#1\right)}
\newcommand{\Sym}[1]{\mathrm{Sym}\left(#1\right)}

\newcommand{\Tr}[1]{\mathrm{Tr}\left[#1\right]}
\newcommand{\Csca}{C_\mathrm{sca}}
\newcommand{\Cabs}{C_\mathrm{abs}}
\newcommand{\Cext}{C_\mathrm{ext}}
\newcommand{\Cpr}{C_\mathrm{pr}}
\newcommand{\musca}{\mu_\mathrm{sca}}
\newcommand{\ave}[1]{\langle #1\rangle}

\newcommand{\microns}{$\mu$m}

\title{Universal Theory of Light Scattering of Randomly Oriented Particles: A Fluctuational-Electrodynamics Approach for Modeling of Light Transport in Disordered Nanostructures}

\author[1,2]{Francisco V. Ramirez-Cuevas}
\author[1]{Kargal L. Gurunatha}
\author[3]{Ivan P. Parkin}
\author[1,*]{Ioannis Papakonstantinou}
\affil[1]{Photonic Innovations lab, Department of Electronic and Electrical Engineering, University College London, London, WC1E 7JE, United Kingdom}
\affil[2]{Facultad de Ingenier\'ia y Ciencias, Universidad Adolfo Ib\'a\~nez, Santiago, Chile}
\affil[3]{Department of Chemistry, University College London, London, WC1H 0AJ, United Kingdom}

\affil[*]{Corresponding Author: i.papakonstantinou@ucl.ac.uk}

\begin{document}

\twocolumn[\begin{@twocolumnfalse}
\maketitle
\begin{abstract}
Disordered nanostructures are commonly encountered in many nanophotonic systems, from colloid dispersions for sensing, to heterostructured photocatalysts. Randomness, however, imposes severe challenges for nanophotonics modeling, often constrained by the irregular geometry of the scatterers involved or the stochastic nature of the problem itself. In this article, we resolve this conundrum by presenting a universal theory of averaged light scattering of randomly oriented objects. Specifically, we derive formulas of \textit{orientation-and-polarization-averaged} absorption cross section, scattering cross section and asymmetry parameter, for single or collection of objects of arbitrary shape, that can be solved by any electromagnetic scattering method. These three parameters can be directly integrated into traditional unpolarized radiative energy transfer modelling, enabling a practical tool to predict multiple scattering and light transport in disordered nanostructured materials. Notably, the formulas of average light scattering can be derived under the principles of fluctuational electrodynamics, allowing analogous mathematical treatment to the methods used in thermal radiation, non-equilibrium electromagnetic forces, and other associated phenomena. The proposed modelling framework is validated against optical measurements of polymer composite films with metal-oxide microcrystals. Our work sets a new paradigm in the theory of light scattering, that may contribute to a better understanding of light-matter interactions in applications such as, plasmonics for sensing and photothermal therapy, photocatalysts for water splitting and CO$_2$ dissociation, photonic glasses for artificial structural colours, diffuse reflectors for radiative cooling, to name just a few.
\end{abstract}
\end{@twocolumnfalse}]

\section{Introduction}
Predicting the complex optical phenomena manifesting in disordered nanomaterials represents a major challenge in the field of computational modeling. 
Plasmonic nanoparticle dispersions for sensing and photothermal therapy\autocite{Langer2020,Cheng2014}, heterostructured photocatalysts for water splitting and CO$_2$ dissociation\autocite{Low2017}, diffuse reflectors for radiative cooling\autocite{Zhao2019}, and porous membranes for solar water desalination\autocite{Tao2018}, are a few examples where the complex interrelation between near-field coupling and multi-scattering interactions with variation in particle morphology, orientation and size impose severe limitations to theoretically predict the system optical response.
Conventional modeling based on computational electromagnetics, such as the Finite Difference or Finite Elements methods, are often unsuitable to quantify the macroscopic optical properties of random media ---namely their specular and total transmittance/reflectance or the intensity distribution---, all critical to assess the performance of these systems. Stochastic methods appear as the most appropriate alternative to calculate these properties, yet, with few exceptions, their applicability is limited to composite media containing subwavelength structures (effective media approximations)\autocite{Cai2009}, or spherical particles (radiative transfer simulations)\autocite{Jacques1995}.
As a result of the limitations in modeling, the majority of designs in disordered nanophotonic materials are driven by a phenomenological approach, whereby multiple samples are fabricated and tested in an iterative process that is time-consuming and expensive.



Light transport in random media is commonly addressed through the radiative transfer theory\autocite{Ishimaru1978,Tsang2000a}, which describes the propagation of the light specific intensity through a composite medium containing a random distribution of independently scattering particles. It is notable that, in principle, the theory is applicable to arbitrary particle geometries and groups of particles\autocite{Tsang2000a}. In practice, however, radiative transfer modeling is most commonly applied to study light transport in composites with spherical particles\autocite{Hogan2014,Jacques1995}. This is due to the scattering properties of the spherical particles being independent of the direction and polarization of the incident light\autocite{Bohren1998}. In this particular scenario, the solution of the radiative transfer equation (RTE) for unpolarized light requires three parameters: the particle's absorption cross section, $\Cabs$, the scattering cross section, $\Csca$, and the asymmetry parameter, $\musca$\autocite{Stamnes2017,Jacques1995}. The latter is an indicator of the scattering anisotropy and a key element to calculate the angular distribution of scattered fields\autocite{Stamnes2017}. 

\begin{figure*}[t!]\centering
\includegraphics[width=1.0\textwidth]{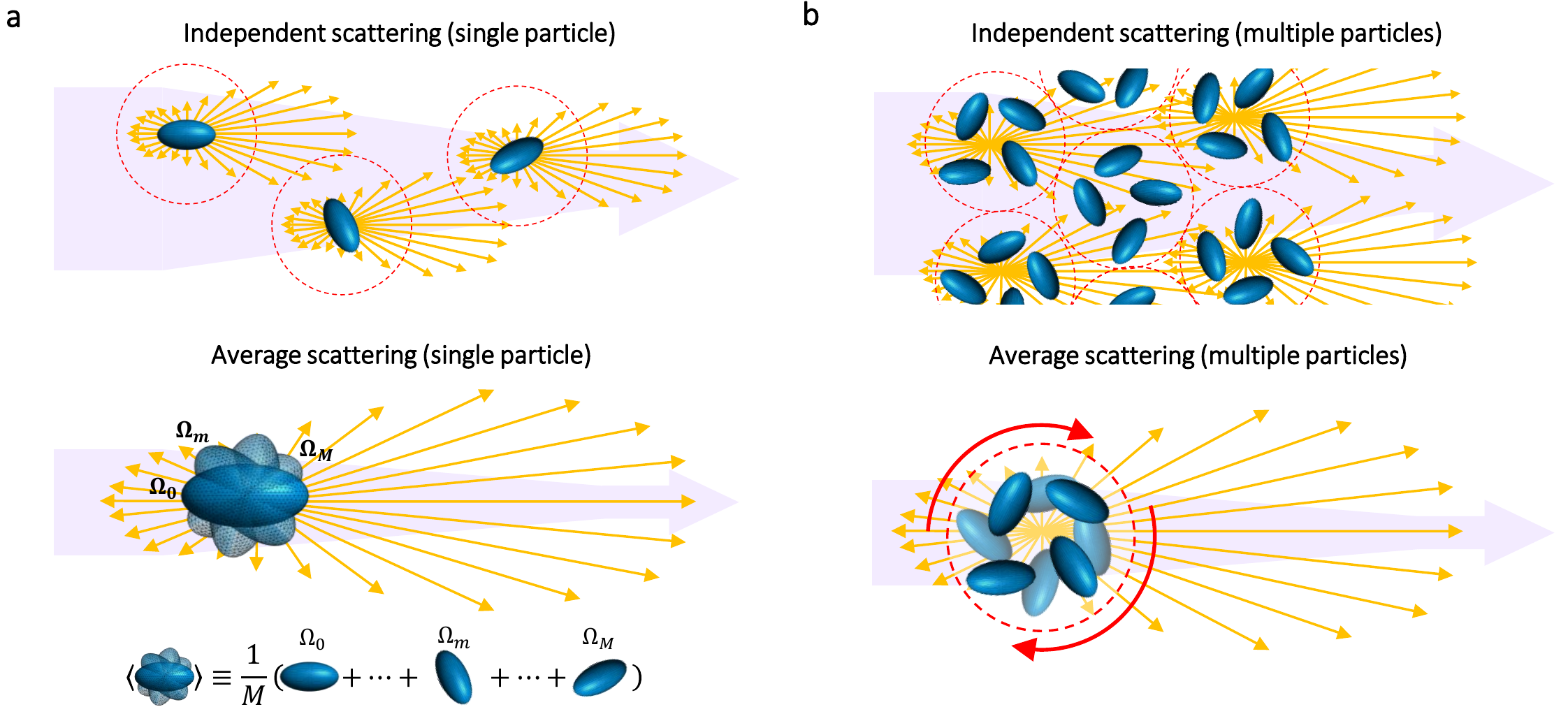}
\caption{\textbf{Average scattering of randomly oriented particles.} \textbf{(a)} A collection of independent scattering particles (in this case prolate spheroids) randomly oriented in space, is approximated by the averaged sum over all possible particle's orientations. In the schematic at the bottom, $\Omega_m$, represents the orientation $m$ of the particle, and $M$ is the total number of particle orientations. In both schematics, the intensity of the incident light beam (purple arrow) decays due to scattering of the particle (yellow arrows). The red dotted circle represents a characteristic domain where short range correlations are relevant. \textbf{(b)} In more dense particle systems, the independent scattering approximation applies to a collection of particles within a properly chosen domain that consider the effects of short range correlations\autocite{Tsang2000b}. Average scattering, thus, is calculated over a characteristic collection of particles.}\label{fig:independent_scattering}
\end{figure*}

The scattering of non-spherical particles, on the other hand, varies with the incident angle and polarization, and the RTE usually becomes too complex to solve\autocite{Tsang2000a}. Under the independent scattering approximation, however, the correlation of the scattered field from different particles vanishes\autocite{Tsang2000a} and the scattering properties of an ensemble of randomly oriented particles can be approximated by the orientation averaged from a single particle [Fig.\ref{fig:independent_scattering}(a)]. For unpolarized light, the RTE becomes scalar \autocite{Hulst1981}, and the \textit{orientation and polarization averaged} $\Cabs$, $\Csca$ and $\musca$ parameter triad can be used for radiative transfer simulations of arbitrary particles, following the same methodology of spherical particles. The same principle could be applied to more complex scenarios; for instance, a medium containing denser particle distributions or even particle clusters. In this case, the averaging should now be performed over a properly chosen collection of particles, for which the effects from short range correlations (due to collective interaction and interference of scattered fields) are prevalent [Fig.\ref{fig:independent_scattering}(b)]\autocite{Tsang2000b}. Similarly, heterogeneous systems with particles of different size and/or optical properties can also be studied.

The computation of orientation and polarization average scattering (average scattering, from now onward) for arbitrary nanoparticles is, however, non-trivial. Standard brute-force methods based on averaging over many plane wave simulations at different angles of incidence and polarizations, can be computationally expensive.\autocite{Suryadharma2018} Alternatively, semi-analytical solutions relying on spherical wave expansion have demonstrated considerable improvements in the efficiency of the calculations\autocite{Bi2013,Mackowski2011,Khlebtsov1992,Suryadharma2018,Fazel-Najafabadi2021}. For example, formulas for direct computation of average scattering have been developed for axially-symmetric objects, such as cylinders\autocite{Bi2013}, spheroids,\autocite{Bi2013} and clusters of spherical particles\autocite{Mackowski2011}. However, the restrictive use of spherical wave basis in this approach still imposes some constrains. Such is the case for objects with no axial symmetry or with sharp edges, where the expansion of the scattered fields into spherical waves is not trivial\autocite{Bi2013}. Often in these problems, the scattered fields are more conveniently expanded through basis relying on surface or volume discretization, such as surface currents in the Boundary Elements Method\autocite{Reid2015} (BEM) or discrete dipoles in the DDA\autocite{Yurkin2007}. For average scattering calculations, nonetheless, the expanded fields have to be transformed into spherical waves, and the efficiency of the method is appreciably reduced\autocite{Yurkin2007}.

In this article, we present a universal theory of average light scattering from randomly oriented scatterers (single or collections of objects) of arbitrary shape, and demonstrate a practical methodology for radiative transfer simulations in disordered nanostructures. The paper is divided into five sections, followed by the conclusions. 
i) In the first section, we derive the formulas for \textit{polarization-and-orientation-averaged} $\Cabs$, $\Csca$ and $\musca$ for arbitrary shaped scatterers. The formulas are independent of the wave basis and, therefore, can be implemented by any computational method of electromagnetic scattering, such as T-Matrix Method\autocite{Tsang2000a}, Boundary elements method (BEM)\autocite{Reid2015}, or DDA\autocite{Yurkin2007}. Based on these results, we evaluate the accuracy limits of other expressions for average light scattering commonly used in the literature\autocite{Santiago2020,Muskens2008}.
ii) In the next section, we demonstrate that the formulas of average light scattering can be derived through the principles of fluctuational electrodynamics, and, hence, can be computed through the mathematical methods used in studies of near-field thermal radiation\autocite{Liu2017a,Cuevas2018,Molesky2019,Kruger2012}, Casimir forces\autocite{Kruger2012} and vacuum friction\autocite{Golyk2013,Mkrtchian2003}. In this context, we develop a computational application to numerically compute averaged light scattering\autocite{scuff-avescatter}, which is based on the fluctuating-surface-current BEM\autocite{Rodriguez2013b,Reid2017}.
iii) In the following section, we validate the theory and simulation code for average scattering simulations against other analytical solutions\autocite{MishchenkoNASA}.
iv) Next, we discuss how the three average light scattering parameters can be applied for modeling of radiative transfer in disordered nanostructures.
v) The accuracy of the modeling framework is demonstrated in the final section, showing excellent agreement with optical measurements of polyethylene (PE) film composites with monoclinic vanadium dioxide [VO$_2$(M)] microcrystals.

\section{Results}

\subsection{Theory of average light scattering of randomly oriented particles}\label{sec:main_theory}

As discussed previously, light scattering from a collection of independent scatterers of arbitrary morphology and randomly oriented in space (Fig.~\ref{fig:independent_scattering}), is equivalent to the average light scattering over all orientations and light polarizations.\autocite{Hulst1981} 
We particularly focus on the average absorption cross section, $\ave{\Cabs}$, scattering cross section, $\ave{\Csca}$, and asymmetry parameter, $\ave{\musca}$, where $\ave{...} = \frac{1}{4\pi}\sum_{P}\int (...)d\Omega$, the $P$ index runs over the two orthogonal polarizations, and $\Omega$ is the solid angle. 
The asymmetry parameter, $\musca$, defines the degree of anisotropy of scattering relative to the direction of the incident light beam, $\ksrc$,\autocite{Bohren1998}:
\begin{equation*}\label{eq:asym_def}
    \musca\Csca =\frac{1}{4\pi}\int_{4\pi}d\kind~ p_\mathrm{sca}(\kind,\ksrc)\kind \cdot \ksrc,
\end{equation*}
\noindent where $\kind$ is the direction of the scattered field and $p_\mathrm{sca}(\kind,\ksrc)$ is the scattering phase function.\autocite{Hulst1981} By definition of $p_\mathrm{sca}$, $\Csca = \frac{1}{4\pi}\int_{4\pi}d\kind~ p_\mathrm{sca}(\kind,\ksrc)$. Thus,  $\musca > 0$ ($\musca < 0$) , represent cases of forward (backward) anisotropic scattering and $\musca = 0$ represents isotropic scattering.

Our derivations are based on the Lippmann-Schwinger approach, a general formalism for electromagnetic scattering phenomena\autocite{Kruger2012} 
In this approach, the scattered fields are given by, $\G_0\T\Esrc$ [in this notation, $\T\Esrc = \int d^3r'~\T(\rpos,\rpos')\cdot\Esrc(\rpos')$], where $\G_0$ is the free space Dyadic Green function and $\T$ is the scattering operator (Supporting Information, Eqs.~(\ref{eq_S:GreenDyadic}) and (\ref{eq_S:T-operator}), respectively). The mathematical form of the operators $\T$ and $\G_0$ is dictated by the expansion basis, and the geometry and optical properties of the scatterer. For example, the $\T$ operator for a spherical particle, expanded in spherical wave basis, is: $\T(\rpos,\rpos') = \sum_l \int d^3r'~\mathbf{f}_l^\mathrm{reg}(\rpos) T_{ll}{\mathbf{f}_l^\mathrm{reg}}^\dagger(\rpos')$; where $\mathbf{f}_l^\mathrm{reg}$ are spherical waves regular at the origin, $T_{ll}$ are the Mie scattering coefficients, and $\dagger$ is the conjugate transpose operator.\autocite{Kruger2012} In numerical methods for electromagnetic scattering, such as the T-Matrix Method\autocite{Tsang2000a}, BEM\autocite{Reid2015}, or DDA\autocite{Yurkin2007}, the form of $\T$ and $\G_0$ has to be computed prior to any scattering calculation.

In this context, the formulas of $\Cabs$ and $\Csca$ for an incident field $\Esrc$ are given by (details in Supporting Information, Section~\ref{sec_S:theory_single}) \autocite{Molesky2020b,Molesky2019}:
\begin{subequations}
\begin{align}
    \Cabs &= \frac{1}{k_0|E_0|^2}\Tr{\left(\Esrc\otimes{\Esrc}^\dagger\right)\T^\dagger~\Asym{- \V^{-1}}\T} \label{eq:Cabs_planewave}
\\
    \Csca &= \frac{1}{k_0|E_0|^2}\Tr{\left(\Esrc\otimes{\Esrc}^\dagger\right)\T^\dagger~\Asym{\G_0}\T}, \label{eq:Csca_planewave}
\end{align}
\end{subequations}
\noindent where, $k_0$ is the wavevector in free space, $E_0$ is the amplitude of the incident fields, $\otimes$ is the tensor product, and $\V$ is the particle's induced potential [Supporting Information, Eq.~(\ref{eq_S:V_potential})], with $\V^{-1} = \G_0 - \T^{-1}$. The operators $\Sym{\A} = (\A + \A^\dagger)/2$ and $\Asym{\A} = (\A - \A^\dagger)/2i$ ---where $\A^\dagger$ is the adjoin of $\A$---, represent the Hermitian and Anti-Hermitian part of $\A$, respectively. The trace is defined as $\Tr{\A} = \sum_l \int d^3r~\A_{ll}(\mathbf{r},\mathbf{r})$.

The formula of $\musca$ is, to the best of our knowledge, presented here for the first time:
\begin{equation}
\begin{split}
\musca = -&\frac{1}{k_0|E_0|^2\Csca}\cdot \\
      &\sum_j\Tr{\ksrc_j\left(\Esrc\otimes{\Esrc}^\dagger\right)\T^\dagger\Sym{\partial_j\G_0}\T},
\end{split}\tag{1c}\label{eq:musca_planewave}
\end{equation}
where the index $j$ in $\ksrc_j$ and in the partial derivative $\partial_j$, represents the global coordinates of the system (\textit{e.g.}, $j = x,y,z$). The formula is derived from the Lorentz force of the scattered fields over the induced currents in an object [Supporting Information, Eq.~(\ref{eq_S:asym_single})].

Because the operators $\T$ and $\G_0$ are dependent on the morphology and optical properties of the scatterer and not its orientation, $\ave{\Cabs}$ and $\ave{\Csca}$ are uniquely determined by the expectation value of the incident fields, $\ave{\Esrc\otimes{\Esrc}^\dagger}$, also known as the free space self-correlator\autocite{Kruger2012}. Explicitly, this term is given by [see derivation in Supporting Information, Eq.~(\ref{eq_S:ave_EE})]:
\begin{equation}\label{eq:ave_EE}
    \ave{\Esrc\otimes{\Esrc}^\dagger} = |E_0|^2\frac{2\pi}{k_0}\Asym{\G_0}.
\end{equation}
\noindent Similarly, $\ave{\musca}$ requires [see derivation in Supporting Information, Eq.~(\ref{eq_S:ave_kEE})]:
\begin{equation}\label{eq:ave_kEE}
    \ave{\ksrc_j(\Esrc\otimes{\Esrc}^\dagger)} = -|E_0|^2\frac{2\pi}{k_0}\Sym{\partial_j\G_0}.
\end{equation}
\noindent 


Using Eqs.~(\ref{eq:ave_EE}) and (\ref{eq:ave_kEE}), we derive the following expressions:
\begin{subequations}\label{eq:ave_all}
\begin{align}
\ave{\Cabs} =& \frac{2\pi}{k_0^2} \Tr{\Asym{\G_0}\T^\dagger\Asym{-\V^{-1}}\T}, \label{eq:abs_ave}
  \\
\ave{\Csca} =& \frac{2\pi}{k_0^2} \Tr{\Asym{\G_0}\T^\dagger\Asym{\G_0}\T}, \label{eq:sca_ave}
  \\
  \begin{split}
\langle \musca\rangle =&
  \frac{1}{\ave{\Csca}}\frac{2\pi}{k_0^4} \\
  &\sum_j\Tr{\Sym{\partial_j\G_0}\T^\dagger\Sym{\partial_j\G_0}\T},
  \end{split}
  \label{eq:asym_ave}
\end{align}
\end{subequations}
\noindent which represent universal recipe for average light scattering, compatible with any method for electromagnetic scattering. As an illustrative example, in the Supporting Information we derive the respective formulas for BEM and T-Matrix using these expressions (Sections \ref{sec_S:theory_BEM} and \ref{sec_S:T-matrix}, respectively). 

The relations Eq.~(\ref{eq:abs_ave}), (\ref{eq:sca_ave}) and (\ref{eq:asym_ave}) can be easily generalized for clustered particles and heterogeneous composites containing different types of particles (Supporting Information, Section~\ref{sec_S:theory_multi}). The light scattering parameters for an individual particle particle $n$ in the cluster, i.e., $\ave{\Cabs^n}$, $\ave{\Csca^n}$ and $\ave{\musca^n}$, are also obtained directly from these relations (details in Supporting Information, Section~\ref{sec_S:theory_indi}). 

We finalize this section by discussing a common approximation for $\ave{\Cabs}$ and $\ave{\Csca}$ found in the literature\autocite{Santiago2020,Muskens2008}:
\begin{align*}
    \ave{\Cabs} \approx \frac{1}{3}\left(C_{\mathrm{abs},x} + C_{\mathrm{abs},y} + C_{\mathrm{abs},z}\right) \\
    \ave{\Csca} \approx \frac{1}{3}\left(C_{\mathrm{sca},x} + C_{\mathrm{sca},y} + C_{\mathrm{sca},z}\right)
\end{align*}
\noindent where $C_{\mathrm{abs},j}$ and $C_{\mathrm{sca},j}$ ($j = x,y,z$) are, respectively, the absorption and scattering cross sections for an incident field polarized in the $j-$direction. As demonstrated in the Supporting Information (Section \ref{sec_S:theory_small}), the expression $\ave{\Cabs} \approx \frac{1}{3}\left(C_{\mathrm{abs},x} + C_{\mathrm{abs},y} + C_{\mathrm{abs},z}\right)$, corresponds to a particular case of Eq.~(\ref{eq:abs_ave}) for subwavelength objects, while analogous expression for $\ave{\Csca}$ holds only for small spherical particles.

\begin{figure*}[t]\centering
\includegraphics[width=1.0\textwidth]{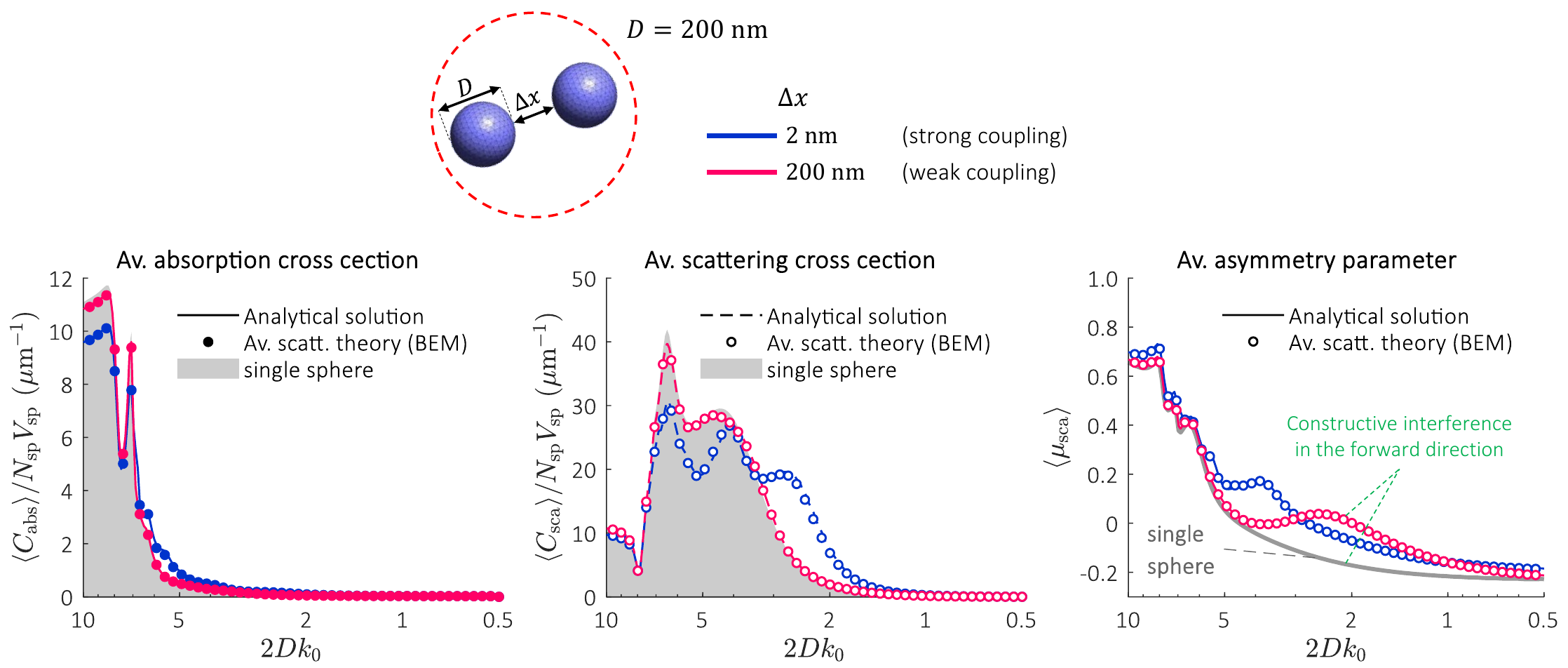}
\caption{\textbf{Average light scattering of randomly oriented silver sphere dimer.} $\ave{\Cabs}$, $\ave{\Csca}$ and $\ave{\musca}$ of the dimer as a function of the size parameter $2Dk_0$ ($k_0=2\pi/\lambda$). The curves $\ave{\Cabs}$ and $\ave{\Csca}$ are normalized to the number of spheres, $N_\mathrm{sp}=2$, and spheres volume, $V_\mathrm{sp}$, for direct comparison with the absorption and scattering cross section of a single sphere. The results are computed by the BEM code for average light scattering simulations\autocite{scuff-avescatter}, and compared against the analytical solution\autocite{MishchenkoNASA}. The scattering parameters of a single sphere  --- i.e., absorption and scattering cross section normalized to the sphere's volume (grey areas), and asymmetry parameter (grey curve) ---, are computed from Mie-scattering \autocite{Bohren1998} and plotted as a reference. The dielectric constant of silver can be found elsewhere\autocite{Babar2015}.}
\label{fig:validation}
\end{figure*}

\subsection{Average light scattering derived from fluctuational electrodynamics}\label{sec:flutuational_electrodynamics}
The trace formulas for $\ave{\Cabs}$, $\ave{\Csca}$ and $\ave{\musca}$ share many similarities with the relations found in studies of fluctuational electrodynamics, namely thermal radiation and non-equilibrium electromagnetic forces\autocite{Kruger2012,Rodriguez2013b}. For example, in the framework of fluctuational electrodynamics, the thermal radiation absorbed by an isolated object, $P_\mathrm{abs}^\mathrm{th}$, is\autocite{Kruger2012}: $$P_\mathrm{abs}^\mathrm{th} = \int_0^\infty d\omega~\Phi_\mathrm{abs}(\omega)\Theta(\omega,T),$$ where 
$\Phi_\mathrm{abs}(\omega) = \frac{2}{\pi}\mathrm{Tr}\{\Asym{\G_0}\T^\dagger\Asym{-\V^{-1}}\T\}$, and $\Theta(\omega,T) = \frac{\hbar\omega}{\exp\left({\hbar\omega/k_\mathrm{B}T}\right) - 1};$ $T$ is the temperature of the environment, $\omega$ is the angular frequency, $k_\mathrm{B}$ is the Boltzmann constant, and $\hbar$ is the reduced Planck constant. 
On the other hand, from light scattering theory\autocite{Bohren1998}, $P_\mathrm{abs}^\mathrm{th} = \int_0^\infty d\omega~4\pi\ave{\Cabs}B_\omega(T),$ where $B_\omega(T) = \frac{k_0^2}{4\pi^3}\frac{\hbar\omega}{\exp\left({\hbar\omega/k_\mathrm{B}T}\right) - 1}$ is the Planck distribution. This leads to the relation:
\begin{equation}\label{eq:AveScat_FDT_relation}
    \ave{\Cabs} = \frac{\pi^2}{k_0^2}\Phi_\mathrm{abs}(\omega).
\end{equation}
\noindent This formula can be easily adapted to compute $\ave{\Csca}$, by replacing $-\V^{-1}$ for $\G_0$.

Another relation comes from the electromagnetic friction induced by a moving photon gas on a stationary object.\autocite{Golyk2013,Mkrtchian2003} To a first order approximation, the friction coefficient $\gamma_f$, can be expressed in terms of $\ave{\Cabs}$, $\ave{\Csca}$ and $\ave{\musca}$ as (Supporting Information, Section \ref{sec_S:vacuum_friction}):
\begin{equation}\label{eq:friction_coef}
    \gamma_f = \frac{\hbar^2k_0^4}{\pi^2k_\mathrm{B}T}\int_0^\infty d\omega 
    \frac{e^{\hbar\omega /k_\mathrm{B}T}}{\left(e^{\hbar\omega/k_\mathrm{B}T} - 1\right)^2}
    \ave{\Cpr}
\end{equation}
\noindent where $\ave{\Cpr} = \ave{\Cext} -  \ave{\musca\Csca}$ and $\ave{\Cext} = \ave{\Cabs} + \ave{\Csca}$, are the average radiation pressure and extinction cross sections, respectively\autocite{Bohren1998}.

As illustrated by Eqs. (\ref{eq:AveScat_FDT_relation}) and (\ref{eq:friction_coef}), the formulas of average scattering can be obtained through the principles of fluctuational electrodynamics. Consequently, the vast library of analytical solutions\autocite{Kruger2012,Golyk2013} and numerical algorithms\autocite{Edalatpour2014,Rodriguez2013b} developed in the context of non-equilibrium energy and momentum transfer can be used to compute $\ave{\Cabs}$, $\ave{\Csca}$ and $\ave{\musca}$. For example, the thermal DDA\autocite{Edalatpour2014} and fluctuating current BEM\autocite{Rodriguez2013b} for thermal radiation simulations, have explicit relations for $\Phi_\mathrm{abs}$ that can be adapted to compute $\ave{\Cabs}$ and $\ave{\Csca}$. 
On the other hand, the fluctuating current BEM also includes routines to compute $\partial\G_0$\autocite{Reid2017}, which can be adapted for $\ave{\musca}$.

In this context, we developed a computational code for average light scattering simulations, based on the fluctuating surface-current BEM formulation for non-equilibrium energy and momentum transfer\autocite{Rodriguez2013b,Reid2017}. The code is implemented as an application of the SCUFF-EM software\autocite{SCUFF-EM}, and can be acceded here\autocite{scuff-avescatter}. Similar to other simulations tools based on BEM\autocite{Rodriguez2013b,Reid2017,Solis2014a,Solis2017}, our code supports objects of arbitrary morphology and groups of objects (Supporting Information, Fig.~\ref{fig_S:BEM_example}), offering a convenient platform to explore the full potential of the average scattering theory presented here.

\begin{figure*}[t!]\centering
\includegraphics[width=1.0\textwidth]{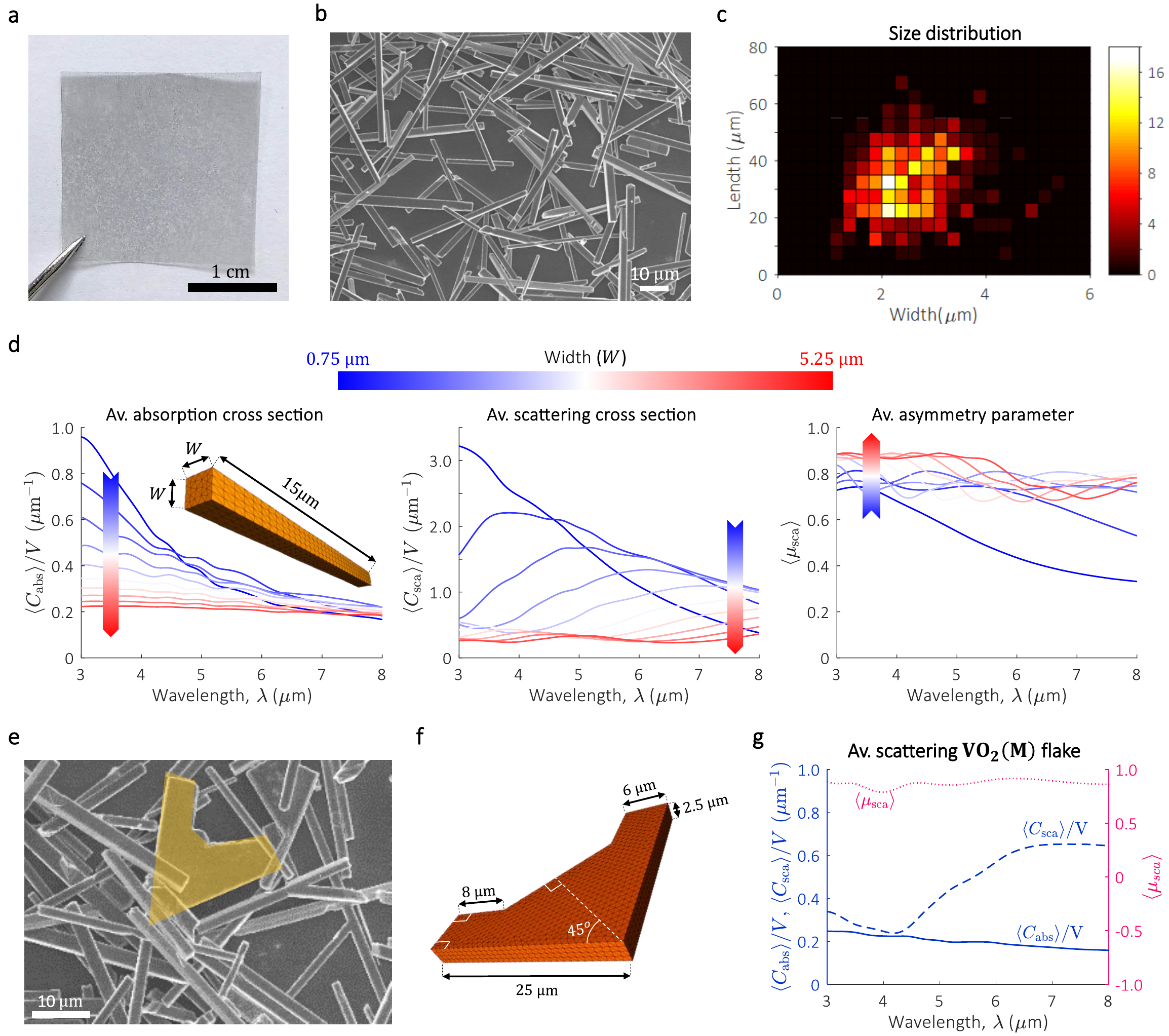}
\caption{\textbf{Characterization and average scattering simulations of Monoclinic Vanadium Dioxide [VO$_2$(M)] microcrystals embedded into a polyethylene (PE) matrix.} \textbf{(a)} Photograph of VO$_2$(M)/PE composite film. \textbf{(b)} SEM of as-grown VO$_2$(M) crystals, which is mainly composed of VO$_2$(M) bars. \textbf{(c)} Size distribution of VO$_2$ bars. \textbf{(d)} $\ave{\Cabs}$, $\ave{\Csca}$ and $\ave{\musca}$ of VO$_2$(M) bars of fixed length, $L = 15$ \microns, and variable width, $W = 0.75-5.25$ \microns~in steps of $0.5$ \microns. The refractive index of the host, $n_\mathrm{host} = 1.5$. The average scattering parameters showed similar dependence to $W$ for other values of $L$ (not shown here). The legend is given by the color bar at the top of the curves. \textbf{(e)} An example of one of the VO$_2$(M) crystals with flake morphology found in the characteristic sample. \textbf{(f)} Computational representation of the VO$_2$(M) flakes, which was considered for the average scattering simulations. \textbf{(g)} Simulated average light scattering of the VO$_2$(M) flake. 
For all average scattering simulations, the refractive index of the host, $n_\mathrm{host} = 1.5$, and the refractive index of the VO$_2$(M) bars was obtained from the literature (see "film2" in Wan et al\autocite{Wan2019}). 
}
\label{fig:PE_VO2_composite}
\end{figure*}

\begin{figure*}[t!]\centering
\includegraphics[width=1.0\textwidth]{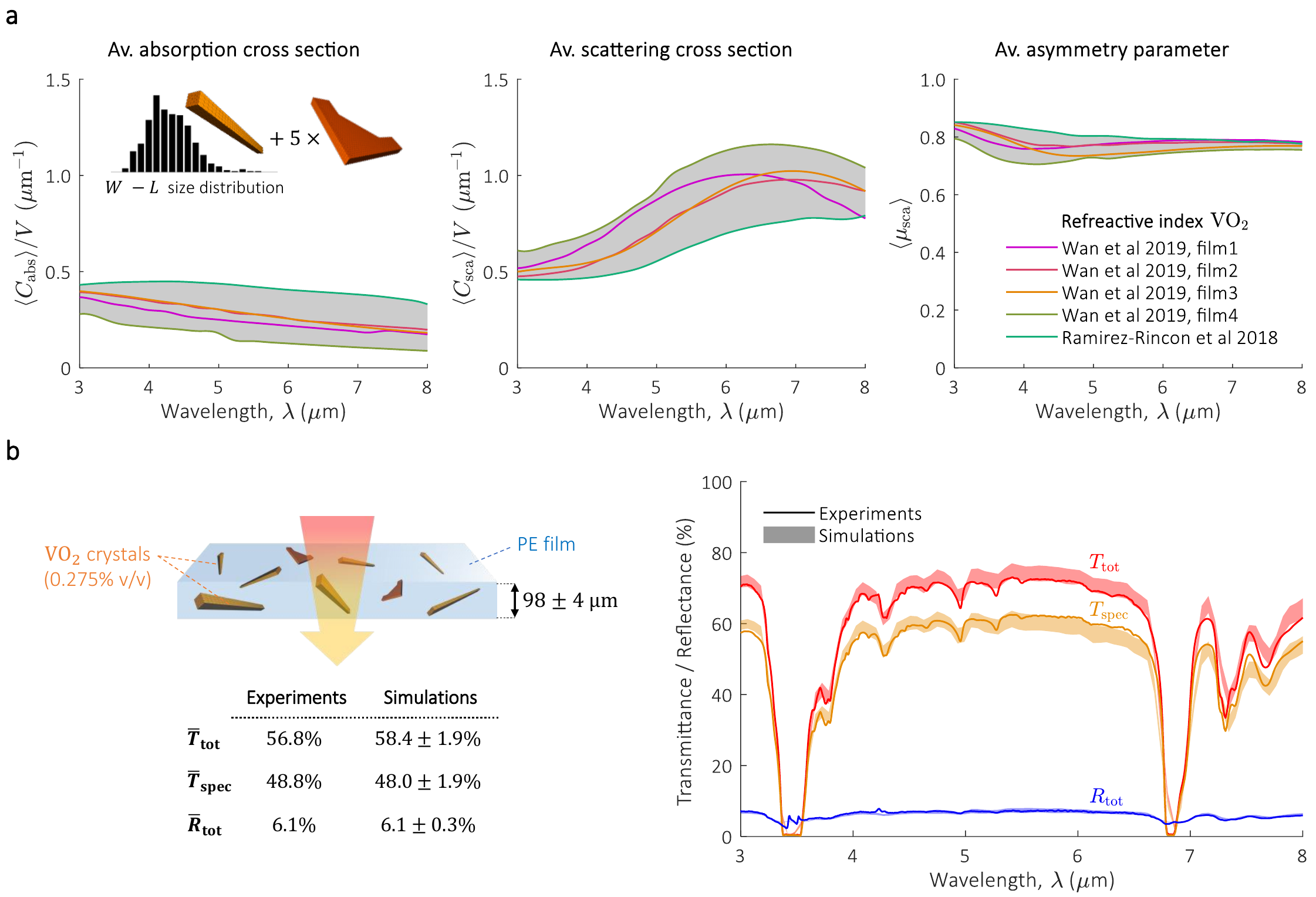}
\caption{\textbf{Radiative transfer modeling of VO$_2$(M)/PE film composite.} \textbf{(a)} $\ave{\Cabs}$, $\ave{\Csca}$ and $\ave{\musca}$ of VO$_2$(M) microcrystal ensemble calculated for five different refractive indexes of VO$_2$(M), as reported by Wan et al 2019\autocite{Wan2019} (labeled as "film1", "film2", "film3" and "film4"), and Ramirez-Rincon et al 2018\autocite{Ramirez-Rincon2018}.
The grey areas mark the upper and lower limit due to variations in the refractive index. For a given refractive index, the curves $\ave{\Cabs}$, $\ave{\Csca}$ and $\ave{\musca}$ were obtained as indicated by the schematic (left figure, inset), i.e. average scattering simulations of bars weighted by the size distribution + average scattering of 5 flakes. Further details in the Supporting Information, Section~\ref{sec_S:ave_scatter_VO2_ensemble}. The curves $\ave{\Cabs}$ and $\ave{\Csca}$ are normalized to the volume of the ensemble $V$ [Supporting Information, Eq.~(\ref{eq_S:Vol_ensemble})]. \textbf{(b)} Validation of radiative transfer theory, showing $T_\mathrm{tot}$, $T_\mathrm{spec}$ and $R_\mathrm{tot}$ of a VO$_2$(M)/PE composite film, as obtained from optical measurements (solid lines) and simulations (filled areas). The optical properties of the PE film were extracted from optical measurements on a clear film (see Supplement, Section~\ref{sec_S:PE_film_nk}). For the simulations, the absorption of PE is considered through the extinction coefficient $\kappa_\mathrm{host}$ (see Methods). The composite is based on a $98\pm4$~\microns~thick PE film with 0.275\% v/v of VO$_2$(M) microcrystals. The upper(lower) limit in the filled areas are the results of variations in the refractive index of VO$_2$(M) and thickness of the film (98-4 or 98+4 \microns).}
\label{fig:PE_VO2_validation}
\end{figure*}

\subsection{Validation of average light scattering theory against analytical solutions}\label{sec:sphere_dimer}

To validate the average scattering theory and BEM simulation code, we consider the problem of light scattering by a randomly oriented sphere dimer (Fig.~\ref{fig:validation}), which has a known solution under the T-matrix approach\autocite{MishchenkoNASA}. The dimer consists of two silver spheres of diameter, $D = 200$ nm, separated by a gap of i) $\Delta x = 2$ nm and ii) $\Delta x = 200$ nm. 
The light scattering parameters of a single sphere obtained from Mie Scattering Theory\autocite{Bohren1998}, are also plotted as a reference.

The results from the average scattering theory show excellent agreement with the analytical solution by Mishchenko et al.\autocite{MishchenkoNASA} (Fig.~\ref{fig:validation}).
When $\Delta x = 2$ nm, the effects of electromagnetic coupling dominate and the average scattering curves of the dimer largely deviate from the response of a single sphere. When $\Delta x = 200$ nm, the coupling effects weaken and both $\ave{\Cabs}$ and $\ave{\Csca}$ approach to the response of a single sphere. However, this is not the case for $\ave{\musca}$. Similar to the optical phenomenon observed in dilute particle media\autocite{Hecht2016Optics}, the scattered fields interfere constructively in the forward direction, which explains the discrepancy between the $\ave{\musca}$ curves.  

As a second test, we consider the problems of average scattering from randomly oriented oblate and prolate spheroids, which has a known solution under the T-matrix approach\autocite{MishchenkoNASA}. The results are displayed in the Supporting Information (Fig.~\ref{fig_S:validation}), showing excellent agreement up to $L_\mathrm{max}/\lambda \gtrsim2.5$, where $L_\mathrm{max}$ is the longest ellipsoid axis. At shorter wavelengths, there is a discrepancy of $\sim5$\% associated to the size of the mesh used in BEM simulations. Even with this discrepancy, the results are consistent with the predictions from the analytical solution, allowing to validate the theory presented here.

\subsection{Radiative transfer modeling for random media with scatterers of arbitrary morphology}\label{sec:light_transport}


For unpolarized light and under the independent scattering approximation, the steady-state RTE for randomly oriented scatterers in a non-absorbing host is\autocite{Tsang2000a}:
\begin{equation}\label{eq:RadTransferEq}
\begin{split}
    \khat\cdot\nabla_\rpos I_\lambda(\rpos,\khat) = 
    &- \frac{f_v}{V_p}\ave{\Cext} I_\lambda(\rpos,\khat) \\
    &+ \frac{f_v}{V_p}\int_{4\pi}d\khat'~\ave{p_\mathrm{sca}(\cos\theta)}I_\lambda(\rpos,\khat'),
\end{split}
\end{equation}
\noindent where $I_\lambda$ is the specific radiative intensity (defined as the energy flux per unit solid angle), $f_v$ is the volume fraction, $V_p$ is the effective volume of the scatterers, $\khat\cdot\nabla_\rpos I_\lambda(\rpos,\khat)$ is the rate of change of $I_\lambda$ at the position, $\rpos$, and direction $\hat{\mathbf{k}}$; and $\ave{p_\mathrm{sca}(\cos\theta)}$ is the orientation and polarization averaged scattering phase function, where $\cos\theta = \khat\cdot\khat'$.  

Commonly, solutions of Eq.~(\ref{eq:RadTransferEq}) consider approximated expressions for the phase function in terms of $\musca$.\autocite{Stamnes2017} For example, the Henyey-Greenstein model:\autocite{Stamnes2017}
\begin{equation}\label{eq:Henyey_Greenstein}
\ave{p_\mathrm{sca}} = \ave{\Csca}\frac{1 - \ave{\musca}^2}{\left[1 + \ave{\musca}^2 - 2\ave{\musca}\cos\theta\right]^{3/2}},
\end{equation}
\noindent is widely used in simulations methods, such as Monte-Carlo,\autocite{Jacques1995} Adding-doubling\autocite{Prahl1993} and Discrete Ordinate\autocite{Stamnes2017}.

As evidenced by Eqs.~(\ref{eq:RadTransferEq}) and (\ref{eq:Henyey_Greenstein}), the RTE and the average light scattering parameters $\ave{\Cabs}$, $\ave{\Csca}$ and $\ave{\musca}$ constitute a complete set to model radiative transfer in composites with scatterers of arbitrary morphology. As demonstrated in the next section, this modeling framework enables to quantitatively predict the macroscopic radiative properties of a composite, such as the total transmittance ($T_\mathrm{tot}$), specular transmittance ($T_\mathrm{spec}$), and total reflectance ($R_\mathrm{tot}$).

\subsection{Validation of radiative transport simulation against experiments}\label{sec:vo2_PE_film}

We demonstrate the accuracy of the previously discussed modeling framework, by comparing the radiative transfer simulations against optical measurements of a composite based on VO$_2$(M) microcrystals embedded in a polyethylene (PE) matrix [Fig.~\ref{fig:PE_VO2_composite}(a)]. We considered VO$_2$(M) microcrystals given its well-defined and highly anisotropic morphology [Fig.~\ref{fig:PE_VO2_composite}(b)], providing an ideal scenario to validate the theory of average scattering and radiative transfer modeling. Additionally, the refractive index \autocite{Wan2019,Ramirez-Rincon2018} and size of microcrystals, ensures a significant contribution from both absorption and scattering in the mid infrared (IR) spectrum\autocite{Bohren1998}.

First, we derived the average light scattering parameters of the VO$_2$(M) microcrystals ensemble using a characteristic sample [Supporting Information, Fig.~\ref{fig_S:VO2M_characterization}(c)]. The size distribution of the bars length ($L$) and width ($W$) is shown in Fig.~\ref{fig:PE_VO2_composite}(c), which assumes bars of square cross section. We calculated $\ave{\Cabs}$, $\ave{\Csca}$ and $\ave{\musca}$ of single VO$_2$(M) bars for $\lambda = 3 - 8$ \microns, considering the range of $W$ and $L$ dictated by the size distribution. The spectrum $\lambda > 8$ \microns~is excluded in the simulations due to the large uncertainty in the refractive index of VO$_2$(M), which is strongly conditioned by crystal orientation, growth method, strain and partial oxidation\autocite{Wan2019}. As shown in Fig.~\ref{fig:PE_VO2_composite}(d), $\ave{\Cabs}$, $\ave{\Csca}$ and $\ave{\musca}$ are strongly sensitive to $W$. On the other hand, the three parameters are less sensitive to changes in $L$, with negligible variations for $L > 15$ \microns~(Supporting Information, Fig.~\ref{fig_S:scatter_vo2_bar_fixedW}). In addition to the VO$_2$(M) bars, we noted small traces of VO$_2$(M) flakes in the sample, such as the one shown in Fig.~\ref{fig:PE_VO2_composite}(e). These VO$_2$(M) flakes are represented by the computational model in Fig.~\ref{fig:PE_VO2_composite}(f), with the simulated average scattering parameters shown in Fig.~\ref{fig:PE_VO2_composite}(g). 

The parameters $\ave{\Cabs}$, $\ave{\Csca}$ and $\ave{\musca}$ of the VO$_2$(M) microcrystals ensemble [Fig.~\ref{fig:PE_VO2_validation}(a)], were estimated using the average scattering simulations of individual bars and the flake, together with the size distribution. We repeated this procedure for five different refractive indexes reported in the literature~\autocite{Wan2019,Ramirez-Rincon2018}, in order to consider the variations in the optical properties of VO$_2$(M). Using the parameters $\ave{\Cabs}$, $\ave{\Csca}$ and $\ave{\musca}$ of the VO$_2$(M) microcrystals ensemble, together with Monte-Carlo simulations of radiative transfer (see details in Methods), we estimate $T_\mathrm{tot}$, $T_\mathrm{spec}$ and $R_\mathrm{tot}$ of a VO$_2$(M)/PE composite film [Fig.\ref{fig:PE_VO2_validation}(b)]. The results are shown by filled areas, representing the upper and lower limits associated with the variations of the refractive index of VO$_2$(M) and thickness of the film. The optical measurements show excellent agreement with the range predicted by simulations, which is further confirmed by comparing the spectral mean of $T_\mathrm{tot}$, $T_\mathrm{spec}$ and $R_\mathrm{tot}$ [Table in Fig.~\ref{fig:PE_VO2_validation}(b)]. The accuracy of the simulation is further confirmed through a second test, which considers a composite film with double concentration of VO$_2$(M) microcrystals (Supporting Information, Fig.~\ref{fig_S:PE_VO2_composite_extra}). 

\section{Conclusion}
We presented a universal theory to predict the average light scattering from randomly oriented objects with arbitrary shape. The formulas of $\ave{\Cabs}$, $\ave{\Csca}$ and $\ave{\musca}$ can be implemented by any method of electromagnetic scattering. Moreover, because these relations are exclusively defined in terms of the operators $\T$ and $\G_0$, they enable more efficient computation than brute-force methods based on averaging over many plane wave simulations (see demonstration in Supporting Information, Section~\ref{sec_S:brute-force}). The general form of the average scattering formulas also provides a convenient landscape to explore the fundamental limits of scattering in random systems. For example, in analogy to the studies of scattering and absorption bounds\autocite{Molesky2020b,Molesky2019}, the limits of forward or backward scattering of randomly oriented particles can be explored through the asymmetry parameter formula [Eq.~(\ref{eq:asym_ave})].

The demonstrated connection between average light scattering and fluctuational electrodynamics enables to extend the theory to other parameters of interest. For example, a formula for the average scattering of moving objects can be extracted from the theory of electromagnetic friction in objects at relative motion\autocite{Golyk2013}. Alternatively, other expressions can be extracted directly through the self-correlators in Eqs.~(\ref{eq:ave_EE}) and (\ref{eq:ave_kEE}), in a similar fashion than the relations obtained from the fluctuation-dissipation theorem\autocite{Kruger2012,Cuevas2018,Molesky2019} 

The parameters $\ave{\Cabs}$, $\ave{\Csca}$ and $\ave{\musca}$ are also practical for radiative transfer simulations for unpolarized light, enabling accurate prediction of the optical properties of composites with scatterers of arbitrary shape, as demonstrated in the study of VO$_2$(M)/PE composite films. The radiative transfer formula for randomly oriented particles [Eq.~(\ref{eq:RadTransferEq})], can be extended to consider other effects present in the light transport process. For example, the emitted thermal radiation from scatterers can be represented through the term $\frac{f_v}{V_p}\ave{\Cabs} B_\omega(T)$ at the right-hand side of the equation\autocite{Stamnes2017}. Similarly, the absorption of the host can be included through the term $-2k_0\kappa_\mathrm{host}I_\lambda(\rpos,\khat)$ at the right-hand side of Eq.~(\ref{eq:RadTransferEq}). 

The methodology used in the study of VO$_2$(M)/PE composite films can be also applied to other composite media, with either dielectrics\autocite{Jacques1995} or metal scatterers\autocite{Hogan2014}, providing that the distance between particles is large enough to ignore the effects of short range correlations. For more complex problems, such as clustered particles or more dense particle distributions\autocite{Hwang2020}, the methodology can be extended using the formulation for multiple objects (Supporting Information, Section \ref{sec_S:theory_multi}). In this case, $\ave{\Cabs}$, $\ave{\Csca}$ and $\ave{\musca}$ must be obtained from simulations over a properly chosen collection of particles that better represents the effects from short-range correlations. The method, thus, could provide key insights to many problems in disordered nanophotonics, such as the effects of agglomeration into the optical absorption of gold nanostars, or the impact of multiple scattering in the light trapping of heterostructured photocatalysts, as we will discuss in future works.

In summary, the theory of average light scattering for randomly oriented objects establishes the underlying basis for fundamental studies in disordered nanophotonics. The combination with radiative transport theory enables a powerful modeling method to predict the macroscale optical response in random systems, setting new pathways for design and optimization of nanophotonic devices based on composites, synthesized nanostructures on substrates or particles in solution.

\section{Methods}
\subsection{Fabrication and characterization of VO$_2$(M) / PE composite films}
The composite was fabricated by dry mixing of VO$_2$(M) microcrystals with low-density PE (LDPE; 42607, Sigma Aldrich) and Ultra-high-molecular-weight PE (UHMWPE; 429015, Sigma Aldrich) at a weight ratio of VO$_2$(M):LDPE:UHMWPE = 1:40:40. The mixture was then melt-pressed into a film at 200$^o$C. The VO$_2$(M) crystals were produced by hydrothermal synthesis using our previously developed procedure\autocite{Gurunatha2020}. The phase of the crystals was confirmed by X-ray diffraction and Raman spectroscopy [Supporting Information, Fig.~\ref{fig_S:VO2M_characterization}(a) and \ref{fig_S:VO2M_characterization}(b), respectively]. The volume fraction of the VO$_2$(M) microcrystals was estimated from the weight ratio and the densities of VO2(M) (4.230 g/cm$^3$)\autocite{Jain2013}, LDPE (0.925 g/cm$^3$ and UHMWPE (0.940 g/cm$^3$). 
 A micrometer was used to characterize the thickness of the film. The reported thickness corresponds to 5 measurements on different sections of the sample.

\subsection{Optical measurements}
The total and specular transmittance, and total reflectance of the VO$_2$(M)/polyethylene composite film were measured with a Fourier-transform-infrared spectrometer (IRTracer-100, Shimadzu) and a mid-IR Integrating sphere (Pike Technologies).

\subsection{Monte-Carlo simulations of radiative transfer}
Radiative transfer simulations were performed by our own code for Monte-Carlo simulations of unpolarized light. The algorithm consist on simulating the trajectories of many individual photons as they interact with particles and interfaces, until they are either, absorbed by particles or exit the simulation domain. The initial condition of each photon is given by the position and direction of the light source. At each simulation step, the optical path ($\Lambda_\mathrm{photon}$) and fate of a photon is estimated by selecting the shortest path between the particle's scattering ($\Lambda_\mathrm{sca}$) and absorption ($\Lambda_\mathrm{abs}$), the absorption of the host ($\Lambda_\mathrm{host}$), or diffraction ($\Lambda_\mathrm{Fresnel}$), where:
\begin{align*}
  \Lambda_\mathrm{sca} &= -\frac{V_p}{f_v\ave{\Csca}}\ln{\xi}, \\
  \Lambda_\mathrm{abs} &= -\frac{V_p}{f_v\ave{\Cabs}}\ln{\xi}, \\
  \Lambda_\mathrm{host} &= -2k_0\kappa_\mathrm{host}\ln{\xi}, \\
\end{align*}
\noindent and $\xi$ is a random number between 0 and 1; $\Lambda_\mathrm{Fresnel}$ is given by the shortest distance between the photon and an interface. In materials with more than one kind of particle, $\Lambda_\mathrm{abs} = \mathrm{min}\left(\Lambda_\mathrm{abs}^i\right)$ and $\Lambda_\mathrm{sca} = \mathrm{min}\left(\Lambda_\mathrm{sca}^n\right)$, where $\Lambda_\mathrm{abs}^n$ and $\Lambda_\mathrm{sca}^n$ are, respectively, the absorption and scattering path from the particle $n$. 

In the case of diffraction ($\Lambda_\mathrm{photon} = \Lambda_\mathrm{Fresnel}$), a photon is either reflected or transmitted by a random selection, with the probabilities of each event proportional to the respective energy flux defined by Fresnel laws. If the photon is absorbed by a particle ($\Lambda_\mathrm{photon} = \Lambda_\mathrm{abs}$) or the host material ($\Lambda_\mathrm{photon} = \Lambda_\mathrm{host}$), the event is terminated and the simulation continues with a new photon at the initial conditions. For a scattered photon ($\Lambda_\mathrm{photon} = \Lambda_\mathrm{sca}$), the new direction is determined by\autocite{Jacques1995}:
\begin{equation*}\label{Witt-HG}
\cos\theta=
\begin{cases}
\frac{1}{2g} \Big\{1 + g^2 -\left[\frac{1 - g^2}{1 - g + 2g \xi}\right]^2\Big\},
 &\text{if}\; g \neq 0,
\\
2\xi - 1, &\text{if}\; g = 0,
\end{cases}
\end{equation*}
\noindent where $g = \ave{\musca}$.

In all our simulations, we considered a slab with large surface area, in order to represent a 2D problem. As a criteria, we selected the smallest surface area by which no photon escapes through the edges. Two large monitors, above and below the slab, measure the total reflectance and transmittance, respectively. The specular transmittance was measured with a third small monitor (1 nm $\times$ 1 nm) at 1 mm distance below the slab. In all the simulation, we considered 1,000,000 photons per wavelength. For validation of our code, refer to Supporting Information, Section~\ref{sec_S:Monte-Carlo}.

\printbibliography

\section*{Acknowledgements}
The work was carried out under the framework of the H2020 European Research Council (ERC) starting grant IntelGlazing grant no: 679891.

\section*{Author contributions statement}
F.V.R and I.P designed the research.F.V.R developed the theory and simulation codes, performed the simulations, and carried the optical measurements. K.L.G carried synthesis and characterization of VO2 (M) bars, and fabricated the VO$_2$(M)/PE composite films. I.P and I.P.P facilitated the research. All authors contributed to writing and reviewing the paper.

\section*{Additional information}
The authors declare no competing interest

\makeatletter\@input{xx_Supplemental.tex}\makeatother
\end{document}


\maketitle
\tableofcontents

%

\section{Theory of average light scattering from random oriented particles}\label{sec_S:theory}
\subsection{Formulation for a single object}\label{sec_S:theory_single}


In the context of Lippmann-Schwinger approach\autocite{Kruger2012}, the presence of a object in an infinitely extended medium $0$ induces a perturbation to the incident (source) electric fields, $\Esrc$, that results in a scattered (induced) electric field, $\Eind$, in medium $0$. For simplicity, we consider medium $0$ as vacuum, with electrical permittivity, $\varepsilon_0$, and magnetic permeability, $\mu_0$, while the object is characterized by linear electromagnetic properties, i.e., dielectric constant, $\bbespilon$, and relative permeability $\bbmu$, which can be, in general, nonlocal complex tensors. The total electric field, $\Etot$, that results from the superposition of $\Esrc$ and $\Eind$, satisfies the following equation\autocite{Kruger2012}:
\begin{equation}\label{eq_S:helmholtz}
\left[\Ho - k_0^2\I - \V\right]\Etot = 0, 
\end{equation}
\noindent where $k_0 = \omega\sqrt{\varepsilon_0\mu_0}$ is the wavevector in free space, $\mathbb{I}$ is the identity operator, $\Ho = \nabla\times\nabla\times$, and
\begin{equation}\label{eq_S:V_potential}
\V = k_0^2\left(\bbespilon - \I\right) + \nabla\times\left(\I - \bbmu^{-1}\right)\nabla\times
\end{equation}
\noindent is the induced potential by the object, with $\V = 0$ anywhere outside the object.

Solution of Eq.~\eqqref{eq_S:helmholtz} is given by:
\begin{equation}\label{eq_S:general_sol}
    \Etot = \left(\I + \G_0\T\right)\Esrc,
\end{equation}
\noindent where $\G_0$ is the free space dyadic green function\autocite{Tsang2000a}:
\begin{equation}\label{eq_S:GreenDyadic}
    \G_0\left(\rpos,\rpos'\right) = \left[\I + \frac{1}{k_0^2}\nabla\nabla\right]\frac{e^{i\mathbf{k}_0|\rpos - \rpos'|}}{4\pi|\rpos - \rpos'|},
\end{equation}

\noindent which satisfies $\left[\Ho - \I\right]\G_0 = \delta(r - r')$, and $\T$ is the scattering operator defined by\autocite{Molesky2019,Kruger2012}:
\begin{equation}\label{eq_S:T-operator}
\T = \left[\V^{-1}- \G_0\right]^{-1}
\end{equation}
\noindent $\T = 0$ anywhere outside of the object $n$.


We begin our discussion with the definition of the absorption, $\Pabs$, and scattering, $\Psca$, energy fluxes in terms of the $\T$-operator \autocite{Molesky2020b}:
\begin{subequations}\label{eq_S:Pflux_single}
\begin{align}
    \Pabs &= \frac{1}{2k_0Z_0}\Tr{\left(\Esrc\otimes{\Esrc}^\dagger\right)\T^\dagger~\Asym{- \V^{-1}}\T}\label{eq_S:Pabs_single}
\\
    \Psca &= \frac{1}{2k_0Z_0}\Tr{\left(\Esrc\otimes{\Esrc}^\dagger\right)\T^\dagger~\Asym{\G_0}\T}\label{eq_S:Psca_single}
\end{align}
\end{subequations}
\noindent which are derived from energy conservation and the relations\autocite{Molesky2020b}:
\begin{equation}\label{eq_S:ind_relations}
    \Jind = \frac{i}{k_0Z_0}\V\Etot
\quad
\mathrm{and}
\quad
    \Eind = \G_0\T\Esrc
\end{equation}
The absorption and scattering cross sections [Main text, Eqs.~\eqqref{eq:Cabs_planewave} and \eqqref{eq:Csca_planewave}, respectively] are given respectively by\autocite{Bohren1998}, $\Cabs = \Pabs/\frac{|E_0|^2}{2Z_0}$, $\Csca = \Psca/\frac{|E_0|^2}{2Z_0}$.

As discussed in the main text, to compute $\ave{\Cabs}$ and $\ave{\Csca}$, we need an expression for $\ave{\Esrc\otimes{\Esrc}^\dagger}$. For an incident field of the form $\Esrc = E_0 e^{ik_0\ksrc(\rpos - \rpos_0)}$ , where $\ksrc$ is the direction of the plane wave and $\rpos_0$ represent the origin of the field away from the object, and considering the two orthogonal polarization $\hat{s}$ and $\hat{p}$: 
\begin{align*}\label{eq_S:ave_EE}
    \ave{\Esrc\otimes{\Esrc}^\dagger} &= \frac{1}{4\pi}\int_{4\pi} d\Omega ~
    |E_0|^2\left[\hat{s}\hat{s}  + \hat{p}\hat{p}\right] 
    e^{ik_0\ksrc\cdot(\rpos - \rpos')} 
    \\
    &= |E_0|^2\frac{1}{4\pi}
    \int_{4\pi} d\Omega~ 
    \left[\hat{s}\hat{s}  + \hat{p}\hat{p}\right]\frac{1}{2}
    \left[e^{ik_0\ksrc\cdot(\rpos - \rpos')} +  e^{-ik_0\ksrc\cdot(\rpos - \rpos')}\right]
    \\
    &= |E_0|^2
    \frac{1}{4\pi}\int_{4\pi} d\Omega 
    \left[\hat{s}\hat{s}  + \hat{p}\hat{p}\right]
    \frac{1}{k_0^2}
    \int_0^\infty k_r^2dk_r~\frac{1}{2}
    \left[\delta\left(k_r - k_0\right) + \delta\left(k_r + k_0\right)\right]
    e^{ik_r\ksrc\cdot(\rpos - \rpos')} 
    \\
    &= |E_0|^2\frac{2\pi}{k_0}
    \frac{1}{(2\pi)^3}\int_{-\infty}^\infty d^3\mathbf{k}~
    \left[\hat{s}\hat{s}  + \hat{p}\hat{p}\right]
    \imag{\frac{1}{k^2 - k_0^2}}
    e^{i\mathbf{k}\cdot(\rpos - \rpos')} 
    \\
    &= |E_0|^2\frac{2\pi}{k_0}\Asym{
    \frac{1}{(2\pi)^3}\int_{-\infty}^\infty d^3\mathbf{k}~
    \left[\I  - \hat{k}\hat{k}\right]
    \frac{e^{i\mathbf{k}\cdot(\rpos - \rpos')}}{k^2 - k_0^2}}
    \\
    &= |E_0|^2\frac{2\pi}{k_0}\Asym{\G_0},\numberthis
\end{align*}
\noindent where the second step exploits the reciprocity of $e^{ik_0\ksrc\cdot(\rpos - \rpos')}$, and the fourth step considers the identity\autocite{Chew2019} $$\frac{1}{2}\left[\delta\left(k_r - k_0\right) + \delta\left(k_r + k_0\right)\right] = \frac{k_0}{\pi}\mathrm{Im}\left(\frac{1}{k^2 - k_0^2}\right).$$ The penultimate expression corresponds to the plane-wave representation of the Dyadic Green function\autocite{Tsang2000a}. Substitution of Eq.~\eqqref{eq_S:ave_EE} into Eqs.~\eqqref{eq:Cabs_planewave} and \eqqref{eq:Csca_planewave} (Main text), respectively, leads the relations Eqs.~\eqqref{eq:abs_ave} and \eqqref{eq:sca_ave} in the Main text.

The asymmetry parameter is given by the force from scattering fields, $\Find$, through the relation\autocite{Mishchenko2001}:
\begin{equation}\label{eq_S:musca_start}
    \musca\Csca = \frac{\omega}{k_0I_0}\ksrc\cdot\Find
\end{equation}

The projection of the scattering force, $\ksrc\cdot\Find$, expressed in terms of the induced currents is given by\autocite{Kruger2012,Polimeridis2015a}:
\begin{align*}
    \ksrc\cdot\Find &=  
    \ksrc\cdot\frac{1}{2\omega}\mathrm{Im}\langle\Jind,\nabla\Eind\rangle
    \\
    &= - \frac{1}{2\omega k_0Z_0}\sum_{j}\Tr{\ksrc_j\left(\Esrc\otimes{\Esrc}^\dagger\right)\T^\dagger\Sym{\partial_j\G_0}\T},
    \numberthis \label{eq_S:asym_single}
\end{align*}
\noindent where we apply the relations from Eq.~\eqqref{eq_S:ind_relations}.

Eq.~\eqqref{eq_S:asym_single} represents a generalized form of the asymmetry parameter, regardless of the form of the source fields. Similarly to Eqs.~\eqqref{eq_S:Pabs_single} and \eqqref{eq_S:Psca_single}, computation of the orientation average of, $\ksrc\cdot\Find$, requires a direct expression for, $\bigg<\ksrc_j\left(\Esrc\otimes{\Esrc}^\dagger\right)\bigg>$:

\begin{align*}\label{eq_S:ave_kEE}
    \bigg<\ksrc_j\left(\Esrc\otimes{\Esrc}^\dagger\right)\bigg> &=
    \frac{1}{4\pi}\int_{4\pi} d\Omega~ 
     \ksrc_j~\left(\hat{s}\hat{s}  + \hat{p}\hat{p}\right)
     |E_0|^2e^{ik_0\ksrc\cdot(\rpos - \rpos')}
    \\
    &= |E_0|^2\frac{2\pi}{ik_0^2}\partial_j\Asym{\G_0}
    \\
    &= |E_0|^2\frac{2\pi}{ik_0^2}\frac{\partial_j\G_0(\rpos,\rpos') - \partial_j\G_0^\dagger(\rpos',\rpos)}{2i}
    \\
    &= |E_0|^2\frac{2\pi}{ik_0^2}\frac{\partial_j\G_0(\rpos,\rpos') + \partial_j'\G_0^\dagger(\rpos',\rpos)}{2i} 
    \\
    &= - |E_0|^2\frac{2\pi}{k_0^2}\Sym{\partial_j\G_0}, \numberthis
\end{align*}

\noindent where, in the third step, we consider the identity $ \partial_j'\G_0^\dagger(\rpos',\rpos) = - \partial_j\G_0^\dagger(\rpos',\rpos)$\autocite{Tai1994}. Substitution of Eqs.~\eqqref{eq_S:ave_kEE} and \eqqref{eq_S:asym_single} into Eq.~\eqqref{eq_S:musca_start}, together with the relation $\ave{\musca} = \frac{\ave{\musca\Cabs}}{\ave{\Csca}}$ leads to the expression Eq.~\eqqref{eq:asym_ave} in the main text.

The expression Eqs.~\eqqref{eq:abs_ave}, \eqqref{eq:sca_ave} and \eqqref{eq:asym_ave} in the Main text, can be extended for linear, homogeneous and non-absorbing external medium, by replacing $k_0$, for $n_0k_0$, where $n_0$ is the refractive index of the external medium.

\subsection{Surface-current BEM formulation}\label{sec_S:theory_BEM}
Particularly for the BEM, under the surface-current formulation, the fields and currents on an object $n$ are, respectively, represented by the bi-linear expressions\autocite{Rodriguez2013b,Reid2015} $$\boldsymbol\phi_n = \begin{Bmatrix} \mathbf{E} \\ \mathbf{H} \end{Bmatrix}\text{\quad and \quad} \boldsymbol\xi_n = \begin{Bmatrix} \mathbf{K} \\ \mathbf{N} \end{Bmatrix},$$ where $\mathbf{K}$ and $\mathbf{N}$ are the surface electric and magnetic current, respectively. 

Using bi-linear expressions, the electromagnetic field is given by $\boldsymbol\phi_n = \mathbbm{\Gamma}_n\boldsymbol\xi_n$. In this formula,
\begin{equation}\label{eq_S:GreeDyadic_bilinear}
    \mathbbm{\Gamma}_n = 
    ik_n\begin{bmatrix}
    Z_n\G_n & \C_n \\
    -\C_n & \frac{1}{Z_n}\G_n
    \end{bmatrix},
\end{equation}
\noindent where $\C_n = \frac{i}{k_n}\nabla\times\G_n$, and $k_n$ and $Z_n$ are the wavevector and impedance of medium $n$, respectively.

Eq.~\eqqref{eq_S:helmholtz_multi} in its bi-linear form is now expressed as: $$\left[\mathbb{H}_0 - ik_0\mathbf{Z}_0^{-1}\I - \sum_n\V_n\right]\boldsymbol\phi^t = 0, 
$$ where $\mathbb{H}_0 = \begin{bmatrix} 0& \nabla\times \\ - \nabla\times & 0\end{bmatrix}$, $\mathbf{Z}_0^{-1} = \begin{bmatrix} Z_0^{-1}& 0 \\ 0 & Z_0\end{bmatrix}$, and
$\V_n = ik_0\mathbf{Z}_0^{-1}\begin{bmatrix} \bbespilon_n - 1& 0 \\ 0 & \bbmu_n - 1\end{bmatrix}$. Notice that in this case, the elements of $\VMult$ and $\GMult^0$, are respectively given by $\V_{nm} = \delta_{nm}\mathbbm{\Gamma}_n ^{-1}$, and $G^0_{nm} = \mathbbm{\Gamma}_0(\rpos_n,\rpos_m)$

The solution is obtained by expanding the surface currents on a particular basis, $\mathbf{f}_r^n(\rpos)$, as: $\xi_n(\rpos) = \sum_r x_r^n \mathbf{f}_r^n(\rpos)$, where the expansion elements $x_r^n$ are obtained from boundary conditions. Further details can be found elsewhere\autocite{Rodriguez2013b,Reid2015}.

From the surface current expansion, the operators-based form of Eqs.~\eqqref{eq_S:abs_ave_multi}, \eqqref{eq_S:sca_ave_multi} and \eqqref{eq_S:asym_ave_multi} is now given in matrix notations by:
\begin{subequations}\label{eq_S:ave_C_BEM}
\begin{align}
   \ave{\Cabs} &= \frac{2\pi}{k_0^2}\Tr{\Sym{G^0}W^\dagger~\Sym{- G^\mathrm{in}}W}\label{eq_S:Cabs_BEM}
    \\
    \ave{\Csca} &= \frac{2\pi}{k_0^2}\Tr{\Sym{G^0}W^\dagger~\Sym{G^0}W}\label{eq_S:Csca_BEM}
    \\
   \ave{\musca\Csca} &= \frac{2\pi}{k_0^4}\sum_{l}\Tr{\Asym{\partial_l G^0}W^\dagger~\Asym{\partial_l G^0}W},\label{eq_S:musca_BEM}
\end{align}
\end{subequations}
\noindent where the elements of the $G^0$ and $G^\mathrm{in}$ are respectively given by $G^{0,nm}_{ij} = \ave{\mathbf{f}^n_i, \mathbbm{\Gamma}_0\mathbf{f}^m_j}$ and $G^{\mathrm{in},nm}_{ij} = \delta_{nm}\ave{\mathbf{f}^n_i, \mathbbm{\Gamma}_m\mathbf{f}^m_j}$, and $W = [G^\mathrm{in} - G^0]$. In this notation, the operator $\langle \: , \: \rangle$, denotes the conjugated inner product: $\langle \mathbf{u}, \mathbf{v}\rangle = \int d^3r~\mathbf{u}^\dagger(\rpos)\cdot\mathbf{v}(\rpos)$, where $\mathbf{u}$ and $\mathbf{v}$ are vector fields. 

Note that due to the imaginary prefactor in Eq.~\eqqref{eq_S:GreeDyadic_bilinear}, the operators "$\Sym{}$" and "$\Asym{}$" in Eqs.~\eqqref{eq_S:Cabs_BEM}, \eqqref{eq_S:Csca_BEM} and \eqqref{eq_S:musca_BEM} are in opposite order in comparison with Eqs.~\eqqref{eq_S:abs_ave_multi}, \eqqref{eq_S:sca_ave_multi} and \eqqref{eq_S:asym_ave_multi}.

\begin{figure}[H]\centering
\includegraphics[width=1.0\textwidth]{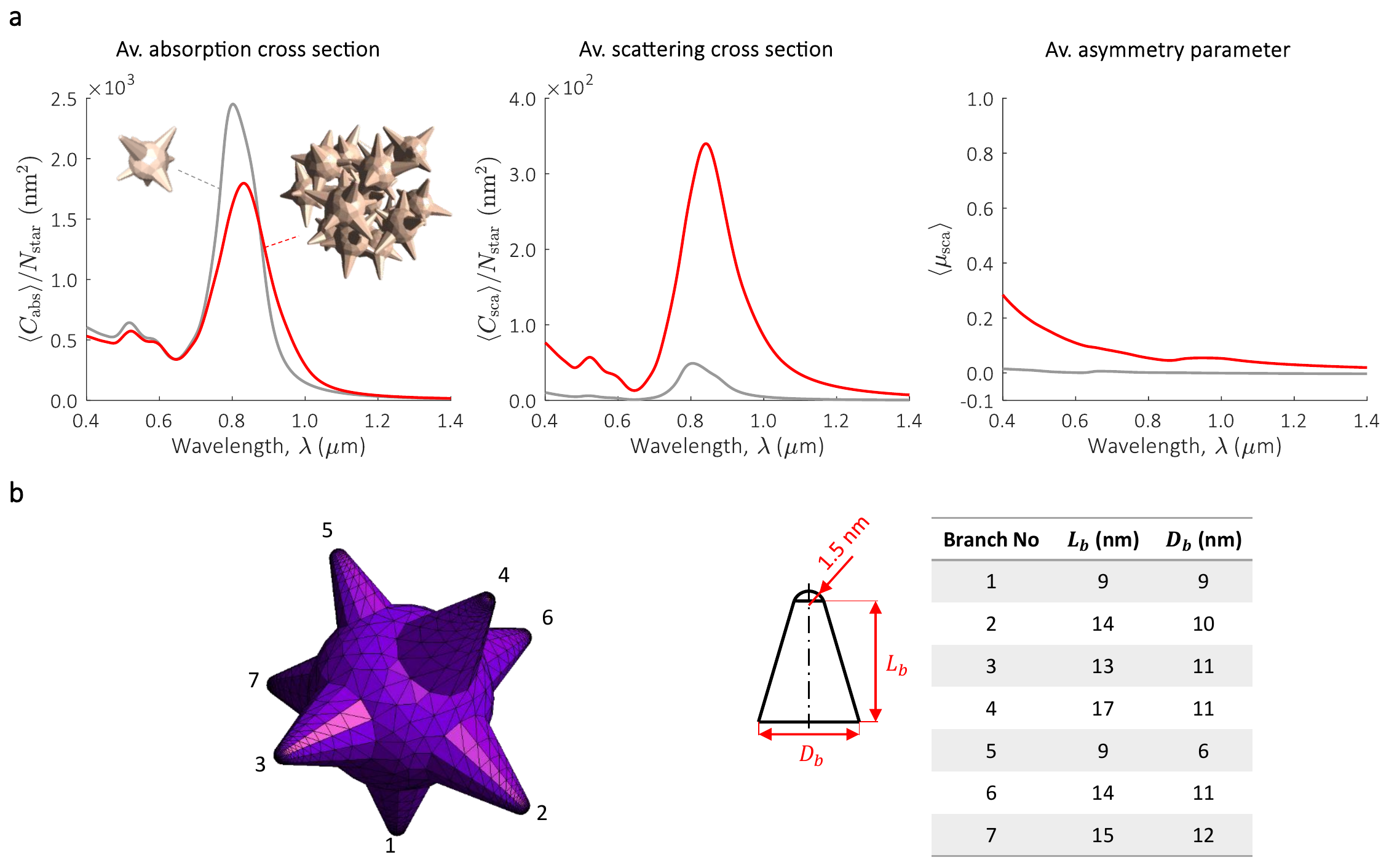}
\caption{\textbf{Average light scattering of agglomerated gold nanostars} \textbf{(a)} $\ave{\Cabs}$, $\ave{\Csca}$ and $\ave{\musca}$ of a single gold nanostar (grey) and cluster of agglomerated gold nanostars (red). The curves $\ave{\Cabs}$ and $\ave{\Csca}$ are normalized to the number of stars ($N_\mathrm{star}$) for direct comparison. The cluster is based on 15 stars at 20\% volume fraction. \textbf{(b)} Model of the star considered for the simulations. The model consist on a star with 7 legs with the dimensions indicated by the table at the right-hand side of the figure. All the simulations were performed by the BEM code for average light scattering simulations\autocite{scuff-avescatter}. The refractive index of gold can be found elsewhere\autocite{Ciesielski2018}.}
\label{fig_S:BEM_example}
\end{figure}

Eqs.~\eqqref{eq_S:Cabs_BEM}, \eqqref{eq_S:Csca_BEM} and \eqqref{eq_S:musca_BEM} were implemented into a BEM application for average light scattering simulations\autocite{scuff-avescatter}. The flexibility of the BEM algorithm enables to study objects of arbitrary morphology.\autocite{Rodriguez2013b} As a demonstration, we simulated the average light scattering parameters $\ave{\Cabs}$, $\ave{\Csca}$ and $\ave{\musca}$ of a single gold nanostar and a cluster of agglomerated gold nanostars (Fig.~\ref{fig_S:BEM_example}). Fig.~\ref{fig_S:BEM_example}(a) illustrates the effects of interparticle coupling and interference of the scattered fields from the stars in cluster, which results in a reduction of $\ave{\Cabs}/N_\mathrm{star}$ and an enhancement of $\ave{\Csca}/N_\mathrm{star}$ in comparison with a single star. Similarly, the scattering anisotropy is also clearly affected in the star cluster, as indicated by the variations in $\ave{\musca}$. Fig.~\ref{fig_S:BEM_example}(b) shows the dimensions of the gold nanostar model considered for the simulations.

\subsection{Average light scattering in Spherical wave basis: T-matrix formulation}\label{sec_S:T-matrix}
The operator $\T$ for spherical wave basis is given by\autocite{Kruger2012}:
\begin{equation}\label{eq_S:T-spherical}
    \T(\rpos,\rpos') = i\sum_{Plm}\sum_{P'l'm'} 
    \Ereg_{Plm}(\rpos) T_{lm,l'm'}^{PP'}\Ereg_{P'l'-m'}(\rpos')
\end{equation}
\noindent where $P = M, N$ correspond to the two orthogonal polarizations, $l = 1,2...\infty$ and $m = -l, ..., 0, ..., +l$; $T_{lm,l'm'}^{PP'}$ are the elements of the T-matrix\autocite{Tsang2000a}, and $\Ereg_{Plm}(\rpos)$ are the spherical waves regular at the origin, defined at the spherical coordinates $r$, $\theta$ and $\phi$\autocite{Kruger2012}:
\begin{align*}
    \Ereg_{Mlm}(\rpos) &= \sqrt{(-1)^m k_0}\frac{1}{\sqrt{l(l+1)}}j_l\left(k_0r\right)\nabla\times Y_l^m(\theta,\phi),
    \\
    \Ereg_{Nlm}(\rpos) &= k_0\nabla\times\Ereg_{Mlm}(\rpos),
\end{align*}
\noindent $j_l$ is the spherical Bessel function of order $l$, and $Y_l^m(\theta,\phi)$ are the spherical harmonics according to the definition from Ref.~\cite{Jackson1998}. Note that ${\mathbf{E}^{\mathrm{reg}*}_{Plm}}(\rpos) = \Ereg_{Nl-m}(\rpos)$, where $^*$ is the complex conjugate\autocite{Kruger2012}.

Because $\G_0$ is a symmetric and reciprocal dyadic, $\Asym{\G_0} = \imag{\G_0}$\autocite{Tai1994}. For spherical wave basis, it can be demonstrated that\autocite{Kruger2012}:
\begin{equation}\label{eq_S:G0-spherical}
    \imag{\G_0} = \sum_{Plm}\Ereg_{Plm}(\rpos)\otimes\Ereg_{Pl-m}(\rpos')
\end{equation}

Replacing Eqs.~\eqref{eq_S:T-spherical} and \eqref{eq_S:G0-spherical} into Eq.~\eqref{eq:abs_ave} and \eqref{eq:sca_ave} (Main text), we derive the expressions of $\ave{\Cabs}$ and $\ave{\Csca}$ for T-matrix, i.e.\autocite{Suryadharma2018,Khlebtsov1992}:
\begin{align}
    \ave{\Cabs} &= \frac{2\pi}{k_0^2}\sum_{Plm}\sum_{P'l'm'}\left[\real{T_{lm,l'm'}^{PP'}}- |T_{lm,l'm'}^{PP'}|^2\right] 
    \label{eq_S:Cabs_T-matrix}
    \\
    \ave{\Csca} &= \frac{2\pi}{k_0^2}\sum_{Plm}\sum_{P'l'm'}|T_{lm,l'm'}^{PP'}|^2
    \label{eq_S:Csca_T-matrix}
\end{align}
\noindent where in the $\ave{\Cabs}$ formula, we consider the relation $\T^\dagger\Asym{-\V^{-1}}\T = \Asym{\T} - \T^\dagger\Asym{\G_0}\T.$

To derive the T-Matrix formula of $\ave{\musca}$, we use the relation: $\Sym{\partial_j\G_0} = \partial_j\imag{\G_0}$; which stems form the reciprocal and anti-symmetric properties of $\partial_j\G_0$\autocite{Tai1994}. For spherical waves basis, this expression is\autocite{Kruger2012}:
\begin{equation}\label{eq_S:G_imag-spherical}
        \partial_j\imag{\G_0} = -\sum_{Plm}\sum_{P'l'm'}
        \mathbf{p_j}^{PP'}_{lm,'l'm'}
        \Ereg_{Plm}(\rpos)\otimes\Ereg_{P'l'-m'}(\rpos')
\end{equation}
\noindent where $\mathbf{p_j}^{PP'}_{lm,'l'm'} = \partial_j\mathcal{V}_{lm,l'm'}^{PP'}(a)|_{a = 0}$, and $\mathcal{V}$ is the translation operator of regular waves\autocite{Rahi2009}. The form of $\mathcal{V}$ for spherical waves is given in Ref.~\cite{Rahi2009}, Appendix C.3.

By replacing Eqs.~\eqref{eq_S:T-spherical} and \eqref{eq_S:G_imag-spherical} into Eq.~\eqref{eq:asym_ave} of the Main text, we derive the T-matrix formulation of $\ave{\musca}$:
\begin{equation}\label{eq_S:musca_T-matrix}
    \ave{\musca} = \frac{1}{\ave{\Csca}}\frac{2\pi}{k_0^2}
    \sum_{j = x,y,z}\sum_{Plm}\sum_{P'l'm'}\sum_{P''l''m''}
    \mathbf{p_j}^{PP'}_{lm,'l'm'}             \; {T^*}_{l'm',l''m''}^{P'P''}
    \mathbf{p_j}^{P''P'''}_{l''m'','l'''m'''} \; T_{l'''m''',lm}^{P'''P}
\end{equation}

For a spherical object, Eq.~\eqref{eq_S:musca_T-matrix} can be simplified to\autocite{Kruger2012,Golyk2013}:
\begin{equation}\label{eq_S:ave-musca_sphere}
    \ave{\musca} = \frac{1}{\ave{\Csca}}\frac{2\pi}{k_0^2}
    \sum_{Plm}\mathrm{Re}\left[
    3a(l,m)^2 T_l^P {T_l^{\Bar{P}*}} + 6b(l,m)^2 T_l^P {T_{l+1}^{P*}}
    \right]
\end{equation}
\noindent where $\Bar{P} = N$ if $P = M$ and vice versa, $T_l^P$ are the mie-scattering coefficients\autocite{Bohren1998}, and:
\begin{align*}
    a(l,m) &= \frac{m}{l(l+1)} \\
    b(l,m) &= \frac{1}{l+1}\sqrt{\frac{l(l+2)(l-m+1)(l+m+1)}{(2l+1)(2l+3)}}
\end{align*}

Using the identities $\sum_{m=-l}^{l}a(n,m)^2 = \frac{2l+1}{3l(l+1)}$ and $ \sum_{m=-l}^{l}b(n,m)^2 = \frac{l(l+2)}{3(l+1)}$, and noting that $\real{T_l^M T_l^{N*}} = \real{T_l^{N} T_l^{M*}}$, we can simplify Eq.~\eqref{eq_S:ave-musca_sphere} to derive the well-known formula\autocite{Bohren1998}:
\begin{equation}
    \ave{\musca} = \frac{1}{\ave{\Csca}}\frac{4\pi}{k_0^2}
    \sum_{l}\mathrm{Re}\left[
    \frac{2l+1}{l(l+1)} T_l^M {T_l^{N*}} + 
    \frac{l(l+2)}{(l+1)} \left(T_l^M T_{l+1}^{M*} + T_l^N T_{l+1}^{N*}\right)
    \right]
\end{equation}

\subsection{Formulation for multiple objects}\label{sec_S:theory_multi}
In the case of multiple objects Eq.~\eqqref{eq_S:helmholtz}, becomes:
\begin{equation}\label{eq_S:helmholtz_multi}
\left[\Ho - k_0^2\I - \sum_{n=1}^{N}\V_n\right]\Etot = 0, 
\end{equation}
\noindent where $N$ is the total number of objects. The solution of this equation is given by:\autocite{Kruger2012} $\Etot = \Esrc + \sum_{n}\G_0\T_n\Esrc_n$, and in matrix form:
\begin{equation}\label{eq_S:general_sol_multi}
    \EtotMult = \EsrcMult + \GMult^0\WMult\EsrcMult
\end{equation}
\noindent where $\EsrcMult$ is a column vector, whose elements contain the incident fields on each object, $\phi_n^\mathrm{i} = \Esrc_n$; $\WMult^{-1} = \GMult_0 -  \VMult^{-1}$ is the analogous of $\T$ for a cluster; $\VMult$ is a band matrix operator whose elements are $\V_{nm} = \V_m\delta_{nm}$; and the matrix operator $\GMult^0$ represents the interaction between the objects in the free space, whose elements are $\G^{0}_{nm} = \G_0\left(\omega;\mathbf{r}_n,\mathbf{r}_m\right)$. The rest of the elements are defined in the main text. 

Following the matrix notation, we extend the relations from Eq.~\eqqref{eq_S:ind_relations} to a vector form in terms of the induced current, $\boldsymbol\xi^\mathrm{s}$, and fields, $\EindMult$:
\begin{equation}\label{eq_S:ind_relations_multi}
        \boldsymbol\xi^\mathrm{s} = -\frac{i}{k_0Z_0}\VMult\EtotMult
\quad
\mathrm{and}
\quad
    \EindMult = \GMult^0\WMult\EsrcMult.
\end{equation}

Analogous to the derivation of $\ave\Cabs$, $\ave\Csca$ and $\ave\musca$ for a single object, we use Eq.~\eqqref{eq_S:general_sol_multi} and the relations in Eq.~\eqqref{eq_S:ind_relations_multi}, to derive:
\begin{subequations}\label{eq_S:ave_all_multi}
\begin{align}
\ave{\Cabs} =& \frac{2\pi}{k_0^2} \Tr{\WMult\,\Asym{\GMult^0}\WMult^\dagger\Asym{-\VMult^{-1}}}, \label{eq_S:abs_ave_multi}
  \\
\ave{\Csca} =& \frac{2\pi}{k_0^2} \Tr{\mathbf{W}\,\Asym{\GMult^0}\WMult^\dagger\Asym{\GMult^0}}, \label{eq_S:sca_ave_multi}
  \\
\langle \musca\rangle =&
  \frac{1}{\ave{\Csca}} \frac{2\pi}{k_0^4}\sum_j\Tr{\WMult\,\Sym{\partial_j\GMult^0}\mathbf{W}^\dagger\Sym{\partial_j\GMult^0}},
 \label{eq_S:asym_ave_multi}
\end{align}
\end{subequations}
\noindent these expressions represent the most general form of the average light scattering theory. They can be applied to single and group of particles, including heterogeneous collections of particles. 

\subsection{Formulation for an individual objects in a cluster}\label{sec_S:theory_indi}
Using Eqs.~\eqqref{eq_S:abs_ave_multi}, \eqqref{eq_S:sca_ave_multi} and \eqqref{eq_S:asym_ave_multi}, we deduce the individual contribution of an object $n$ in the cluster:
\begin{subequations}
\begin{align*}
    \ave{\Cabs^n} = &\frac{2\pi}{k_0^2} \Tr{\Se_{nn}\,\Asym{-\V_n^{-1}}}
    \numberthis\label{eq_S:ave_Cabs_indi}
    \\
    \numberthis\label{eq_S:ave_Csca_indi}
    \begin{split}
    \ave{\Csca^n} = &\frac{2\pi}{k_0^2} \Big\{\Tr{\Se_{nn}\,\Asym{\G_{nn}^0}} 
    + \sum_{n\neq m}^{N}\Tr{\Se_{nm}\,\Asym{\G_{nm}^0}}\Big\}
    \end{split}
    \\\numberthis\label{eq_S:ave_musca_indi}
    \begin{split}
    \ave{\musca^n} = &\frac{N}{\ave{\Csca}}
    \frac{2\pi}{k_0^4}\sum_j\Big\{
    \Tr{\Se^{\nabla}_{j,nn}\,\Sym{\partial_j \G_{nn}^0}} 
    + \sum_{n\neq m}^{N}\Tr{\Se_{j,nm}^{\nabla}\,\Sym{\partial_j \G_{nm}^0}}\Big\},
    \end{split}
\end{align*}
\end{subequations}
\noindent where $\mathbf{S} = \left[\WMult\,\Asym{\GMult^0}\WMult^\dagger\right]$, and $\mathbf{S}^\nabla_j = \left[\WMult\,\Sym{\partial_j\GMult^0}\WMult^\dagger\right]$.

In Eqs.~\eqqref{eq_S:ave_Csca_indi} and \eqqref{eq_S:ave_musca_indi} the first and second term inside the curly brackets represent, self and interference scattering, respectively\autocite{Mishchenko2000}. Note that $\ave{\musca^n}$ in Eq.~\eqqref{eq_S:ave_musca_indi} is scaled by $N$ for better comparison with $\ave{\musca}$ from an isolated object.

To recover the total values of the cluster [Eqs.~\eqqref{eq_S:abs_ave_multi}, \eqqref{eq_S:sca_ave_multi} and \eqqref{eq_S:asym_ave_multi}]:
\begin{align*}
 \ave{\Cabs} &= \sum_{n=1}^N \ave{\Cabs^n} \\
 \ave{\Csca} &= \sum_{n=1}^N \ave{\Csca^n} \\
 \ave{\musca} &= \frac{1}{N}\sum_{n=1}^N \ave{\musca^n}.
\end{align*}


\subsection{Approximation for subwavelength particles}\label{sec_S:theory_small}

For small objects, $\Asym{\G_0} \approx \frac{k_0}{6\pi}\mathbb{I}$\autocite{Novotny2006}, and, $\T \approx 4\pi k_0\pmb{\alpha}$,\footnote{the expression is deducted from the electric field generated by a point dipole\autocite{Jackson1998}} where $\pmb{\alpha}$ is the polarizability tensor. Substitution into Eq.~\eqqref{eq_S:ave_Cabs_indi}, together with the relation $$\T^\dagger\Asym{-\V^{-1}}\T = \Asym{\T} - \T^\dagger\Asym{\G_0}\T,$$ gives:
\begin{equation}\label{eq_S:ave_Cabs_small}
    \ave{\Cabs} = \frac{4\pi}{3}k_0~\Tr{\imag{\pmb{\alpha}}} - \frac{8\pi}{9}k_0^4~\mathrm{Tr}\left[\pmb{\alpha}^\dagger\pmb{\alpha}\right].
\end{equation}

Because the particles are small, the second term in Eq.~\eqqref{eq_S:ave_Cabs_small} is negligible in comparison with the first, leading to $$\ave{\Cabs} \approx \frac{1}{3}\left(C_{\mathrm{abs},x} + C_{\mathrm{abs},y} + C_{\mathrm{abs},z}\right).$$ 

The second term in Eq.~\eqqref{eq_S:ave_Cabs_small} corresponds to the scattering of the particle, which can be written as: $$\ave{\Csca} = \frac{8\pi}{9}k_0^4~\sum_{ij}|\alpha_{ij}|^2,$$ where $\alpha_{ij}$ are the elements of the tensor $\pmb{\alpha}$. If $\pmb{\alpha}$ is a diagonal tensor, $\mathrm{Tr}\left[\pmb{\alpha}^\dagger\pmb{\alpha}\right] = |\alpha_{xx}|^2 + |\alpha_{yy}|^2 + |\alpha_{zz}|^2,$ and we derive the commonly used relation\autocite{Santiago2020,Muskens2008}, $$\ave{\Csca} \approx \frac{1}{3}\left(C_{\mathrm{sca},x} + C_{\mathrm{sca},y}+ C_{\mathrm{sca},z}\right).$$

\begin{figure}[H]\centering
\includegraphics[width=1.0\textwidth]{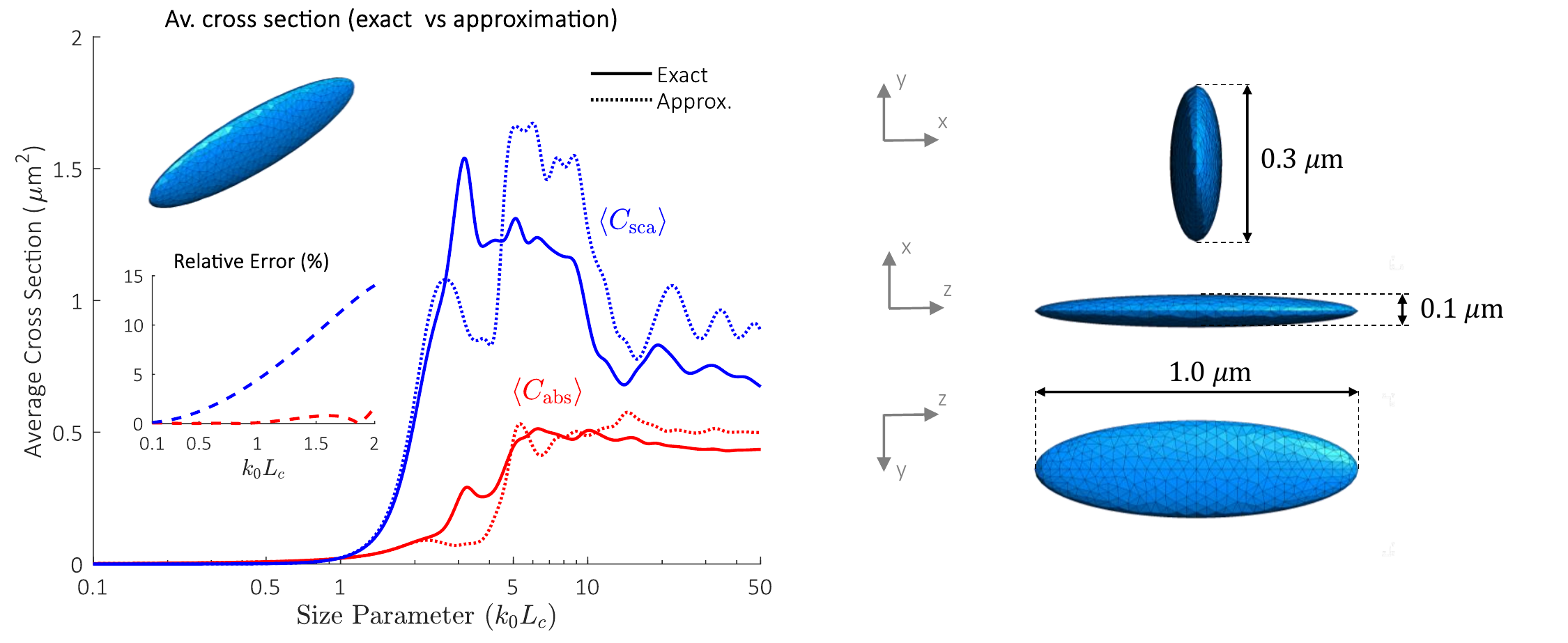}
\caption{\textbf{Average light scattering of randomly oriented particles of subwavelength size. Comparison of exact solution and approximations} The figure shows $\ave{\Cabs}$ and $\ave{\Csca}$ of a randomly oriented ellipsoid plotted against the size parameter $k_0L_c$ ($L_c = 1.0$ \microns). The exact solution for $\ave{\Cabs}$ and $\ave{\Csca}$ is based on Eqs.~\eqqref{eq:abs_ave} and \eqqref{eq:sca_ave} (Main text), respectively, and was computed by our BEM application for average light scattering simulations\autocite{scuff-avescatter}. The approximated results are based on the relations, $\ave{\Cabs} \approx \frac{1}{3}\left(C_{\mathrm{abs},x} + C_{\mathrm{abs},y} + C_{\mathrm{abs},z}\right)$ and $\ave{\Csca} \approx \frac{1}{3}\left(C_{\mathrm{sca},x} + C_{\mathrm{sca},y} + C_{\mathrm{sca},z}\right)$. The inset shows the relative error between the approximated and the exact solutions for small size parameters. The dimensions of the ellipsoid are indicated at the right figure. The refractive index of the ellipsoid is $N = 3.5 + 0.1i$.}
\label{fig_S:aveC_approx}
\end{figure}

In Fig.~\ref{fig_S:aveC_approx}, we compute $\ave{\Cabs}$ and $\ave{\Csca}$ for a randomly oriented ellipsoid, using the exact solution [Main text, Eqs.~\eqqref{eq:abs_ave} and \eqqref{eq:sca_ave}, respectively] and the approximation for small objects. The ellipsoid represents a typical case where $\pmb{\alpha}$ is not diagonal. As shown in the inset of Fig.~\ref{fig_S:aveC_approx}, the approximation $\ave{\Csca} \approx \frac{1}{3}\left(C_{\mathrm{sca},x} + C_{\mathrm{sca},y} + C_{\mathrm{sca},z}\right)$, induces considerable error, even when $k_0L_c$ is small. On the other hand, the approximation for $\ave{\Cabs}$, shows a small relative error ($<2\%$) for small particles ($k_0L_c < 2$). For larger particles ($k_0L_c > 2$), both approximations fail.

\subsection{Efficiency of the theory against brute-force averaging methods.}\label{sec_S:brute-force}

Consider, for example, the computation of $\ave{\Csca}$ by averaging over many plane-wave simulations at different angles of incidence (brute-force averaging). In terms of the Lippmann-Schwinger approach to scattering, this is equivalent to:
\begin{equation}\label{eq_S:brute-force}
    \ave{\Csca} = \frac{1}{k_0|E_0|^2}\frac{1}{M}\sum_m^{M}\Tr{\left(\Esrc_{m}\otimes{\Esrc_{m}}^\dagger\right)\T^\dagger~\Asym{\G_0}\T}
\end{equation}
\noindent where $m$ indicates a plane-wave simulations at a particular angle of incidence and polarization, and $M$ is the total number simulations. This expression is constructed from the definition of $\Csca$ for an incident field $\Esrc$ [Main text, Eq.~(\ref{eq:Csca_planewave})].

Similar to other formulas derived in this work, Eq.~(\ref{eq_S:brute-force}) is a general recipe that illustrates how to compute $\ave{\Csca}$ using brute-force averaging. From this formula, we can deduct the steps required by any method of electromagnetic scattering, i.e.:
\begin{enumerate}
    \item Determine the mathematical form of $\T$ and $\G_0$, for a particular expansion basis.
    \item Compute the expansion of $\Esrc_m$ using the particular basis.
    \item Repeat "step 2" $M$ times.
    \item Replace $\Esrc_m$, $\T$ and $\G_0$ into Eq.~(\ref{eq_S:brute-force})
\end{enumerate}
\noindent As discussed in the Section~\ref{sec_S:theory_small}, $M \ge 3$, where $M = 3$ correspond to the particular case of subwavelength particles.

On the other hand, as illustrated by Eq.~\eqqref{eq:sca_ave} in the Main text, the theory of average light scattering does not require $\Esrc_m$. Consequently, the computation of $\ave{\Csca}$ is reduced to only two steps:
\begin{enumerate}
    \item Determine the mathematical form of $\T$ and $\G_0$, for a particular expansion basis.
    \item Replacing $\T$ and $\G_0$ into Eq.~\eqqref{eq:sca_ave} of the Main text.
\end{enumerate}
\noindent The same arguments apply for $\ave{\Cabs}$ and $\ave{\musca}$. Thus, the theory of average light scattering enables more efficient computation of $\ave{\Cabs}$, $\ave{\Csca}$ and $\ave{\musca}$, than brute-force averaging methods.

\section{Average light scattering from fluctuational electrodynamics}
\subsection{Vacuum friction and averaged light scattering}\label{sec_S:vacuum_friction}
An isolated object moving at a small velocity $\mathbf{v}$ experiences a non-conservative friction force, $\mathbf{F}_f$, that results from the interaction with thermal fluctuations in the free space. To a first order approximation, the vacuum friction is given by, $\mathbf{F}_f = -\hat{\gamma}\cdot\mathbf{v}$, where $\hat{\gamma}$ is the friction tensor given by\autocite{Golyk2013}:
\begin{equation}\label{eq_S:friction_tensor}
    \hat{\gamma}_{ij} = \frac{2\hbar^2}{\pi k_\mathrm{B}T}\int_0^\infty d\omega 
    \frac{e^{\hbar\omega /k_\mathrm{B}T}}{\left(e^{\hbar\omega/k_\mathrm{B}T} - 1\right)^2}
    \mathrm{Im}\{\Tr{\partial_i\left(1 + \G_0\T\right)\partial_j\Asym{\G_0}\T^\dagger}\},
\end{equation}
\noindent here, $k_\mathrm{B}$ is the Boltzmann constant, and $T$ is the equilibrium temperature of the system. Integrating the vacuum friction force over all solid angles leads to the relation\autocite{Mkrtchian2003}, $\ave{\mathbf{F}_f} = \gamma\mathbf{v}$, where $\gamma = \sum_i \hat{\gamma}_{ii}$ is the friction coefficient.

From the relations, $\partial_i\Asym{\G_0} = -i\Sym{\partial_i\G_0}$\autocite{Golyk2013}, and, $\partial_i^2\Asym{\G_0} = - k_0^2\Asym{\G_0}$,\footnote{This expression is based on the fact that $\Asym{\G_0}$ can be expressed by propagating waves, as demonstrated in Eq.~\eqqref{eq_S:ave_EE}} the friction coefficient is expressed as:
\begin{equation}\label{eq_S:friction_coef}
\begin{split}
    \gamma = \frac{2\hbar^2}{\pi k_\mathrm{B}T}\int_0^\infty d\omega 
    \frac{e^{\hbar\omega /k_\mathrm{B}T}}{\left(e^{\hbar\omega/k_\mathrm{B}T} - 1\right)^2}
    &\{
     k_0^2\Tr{\Asym{\G_0}\Asym{\T}} 
     \\
     &- \sum_i\Tr{\Sym{\partial_i\G_0}\T^\dagger~\Sym{\partial_i\G_0}\T}
    \}.
\end{split}
\end{equation}

Applying Eqs.~\eqqref{eq:abs_ave} and \eqqref{eq:sca_ave} (Main text) into the first term inside the curly brackets,\footnote{we consider the relation: $\Asym{\T} = - \T^\dagger\Asym{\T^{-1}}\T = \T^\dagger\Asym{\G_0 - \V^{-1}}\T$} and Eq.~\eqqref{eq:asym_ave} (Main text) into the second, we derive Eq.~\eqqref{eq:friction_coef} from the main text.

\section{Validation of average light scattering theory against analytical solutions}
\subsection{Average light scattering of randomly oriented spheroids}\label{sec_S:theory_validation}

\begin{figure}[H]\centering
\includegraphics[width=1.0\textwidth]{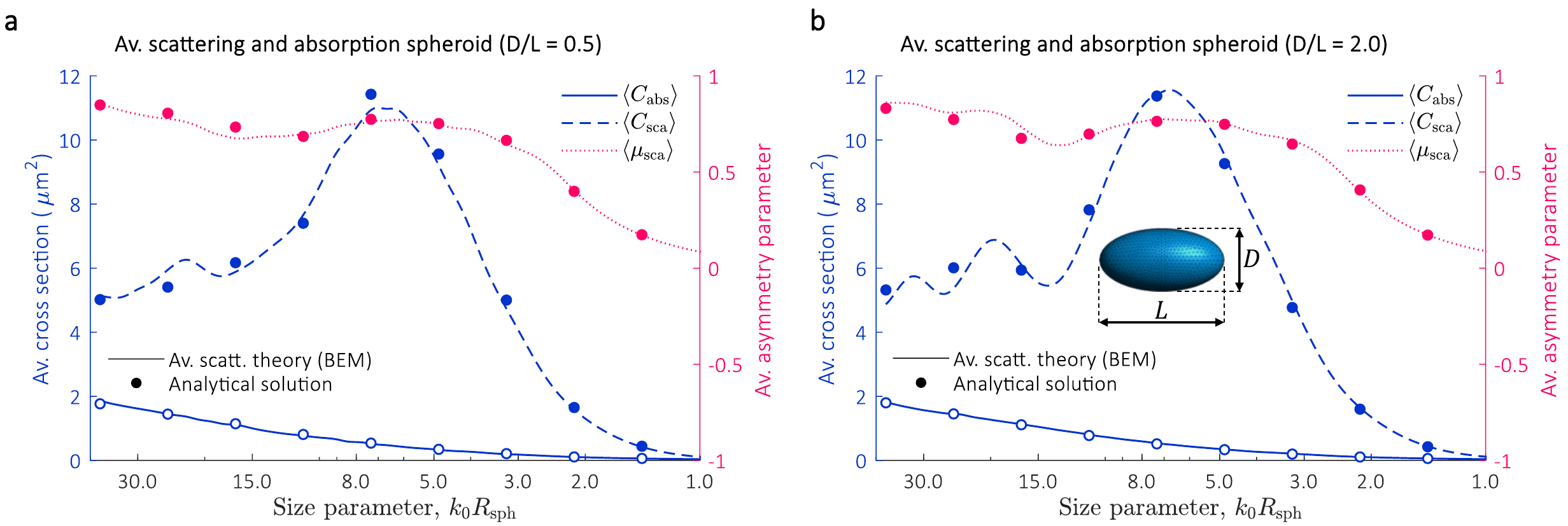}
\caption{\textbf{Average light scattering of randomly oriented spheroids.} $\ave{\Cabs}$, $\ave{\Csca}$ and $\ave{\musca}$ of \textbf{(a)} prolate ($D/L = 0.5$) and \textbf{(b)} oblate spheroids ($D/L = 2.0$), as a function of the size parameter $k_0R_{\mathrm{sph}}$, where $R_{\mathrm{sph}} = 1$ is the radius of a sphere with equivalent surface area\autocite{Mishchenko1996a}. The results are computed by the BEM code for average light scattering simulations\autocite{scuff-avescatter}, and compared with the analytical solutions\autocite{MishchenkoNASA}. The dielectric constant of the ellipsoids is $\varepsilon_p = 2.2499 + 0.0240i$. The principal and secondary axis of the prolate(oblate) spheroid are $L = 1.5304$(0.6016) and $D = 0.7653$(1.2032), respectively.}
\label{fig_S:validation}
\end{figure}

We validate our code for average scattering simulations\autocite{scuff-avescatter} against the analytical solution for prolate [Fig.~\ref{fig_S:validation}(a)] and oblate [Fig.~\ref{fig_S:validation}(b)] spheroids\autocite{MishchenkoNASA}. In terms of relative error, the agreement of the simulations for the prolate(oblate) spheroid is 3.6\%(1.6\%), 4.5\%(3.7\%) and 2.6\%(3.2\%) for $\ave{\Cabs}$, $\ave{\Csca}$ and $\ave{\musca}$, respectively. The discrepancy becomes more significant at short wavelengths, specifically at $L_\mathrm{max}/\lambda > 2.37$($L_\mathrm{max}/\lambda > 2.80$) for the prolate(oblate) spheroid.

\section{Validation of radiative transfer simulation against experiments}\label{sec_S:rad_transfer_sim}

\subsection{Characterization of as-grown VO$_2$(M) microcrystals}\label{sec_S:VO2M_characterization}
\begin{figure}[H]\centering
\includegraphics[width=1.0\textwidth]{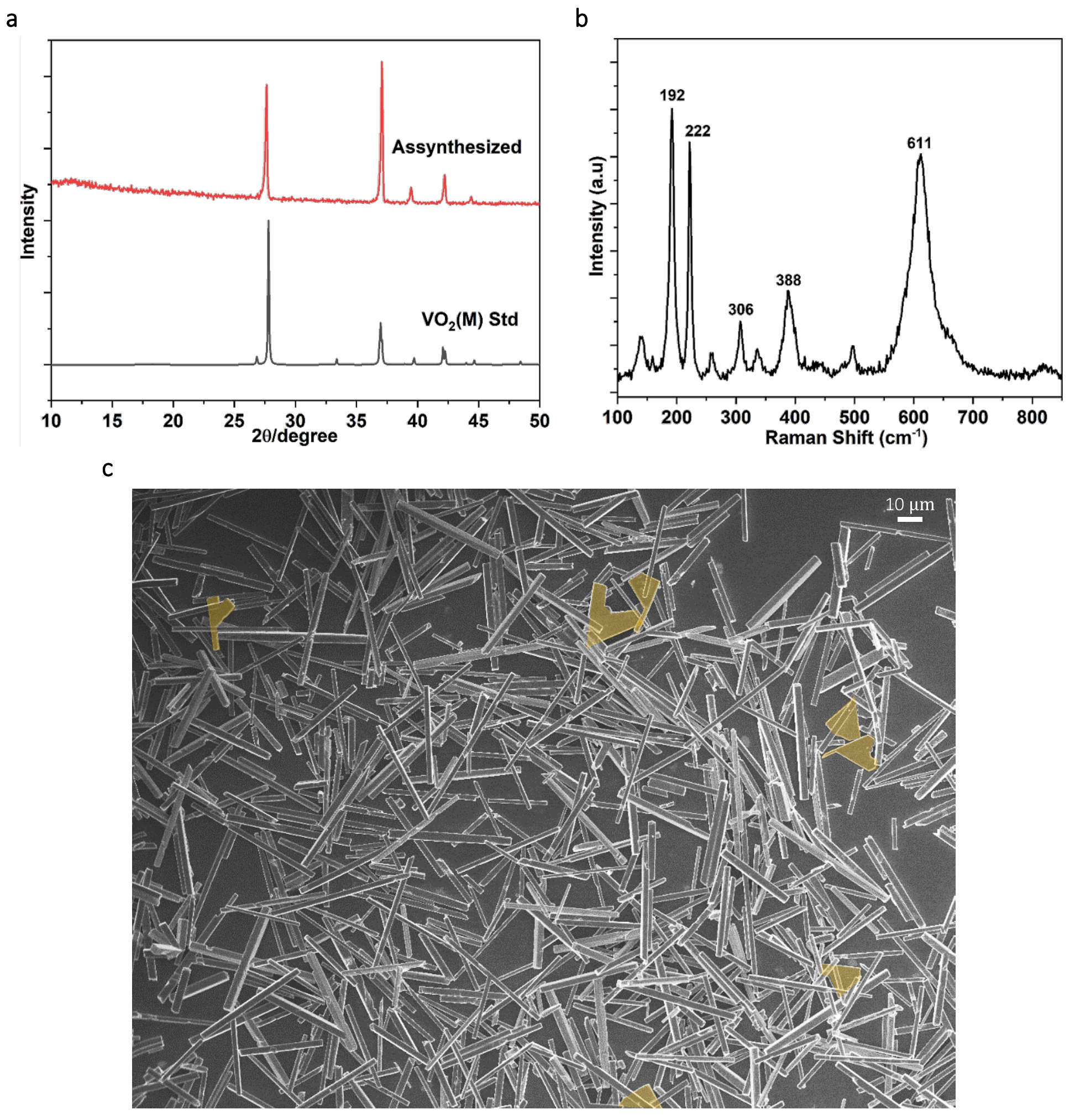}
\caption{\textbf{Characterization of as-grown VO$_2$(M) powder.} \textbf{(a)} X-ray diffraction and \textbf{(b)} Raman shift spectroscopy of VO$_2$(M) microcrystals grown by hydrothermal synthesis (see Main Text, Materials and Methods). \textbf{(c)} SEM image of VO$_2$(M) powder used for the estimation of size distribution [Main text, Fig.~\ref{fig:PE_VO2_composite}(c)]. Most crystals have a bar morphology with few having a larger flat morphology (flakes), which are marked in yellow.}
\label{fig_S:VO2M_characterization}
\end{figure}

\subsection{Average light scattering of VO$_2$(M) bars as a function of $L$.}\label{sec_S:scatter_vo2_bar_fixedW}

\begin{figure}[H]\centering
\includegraphics[width=1.0\textwidth]{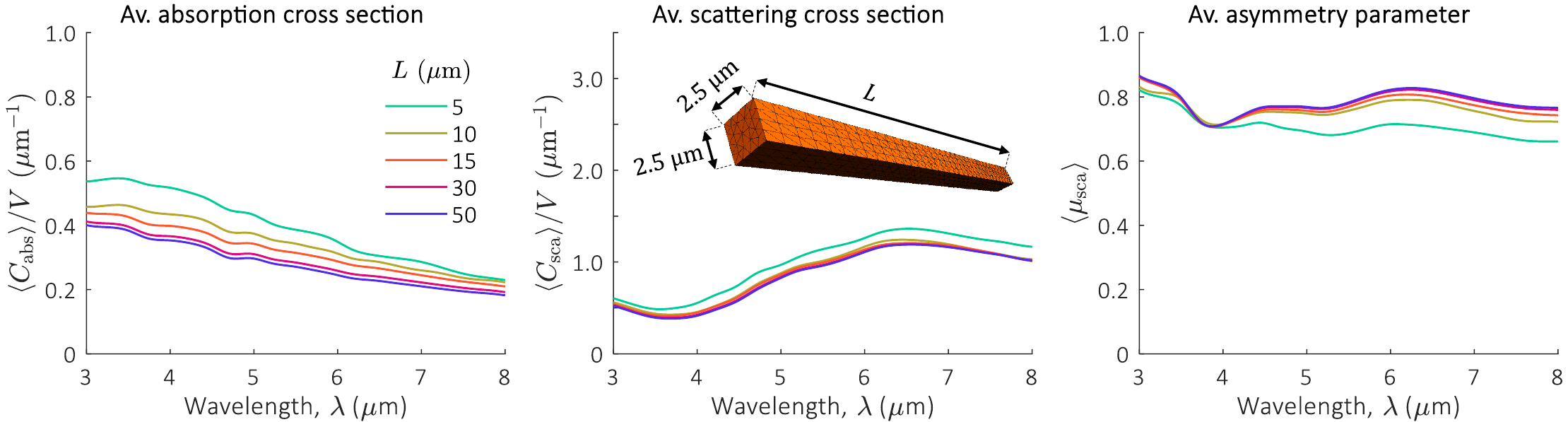}
\caption{\textbf{$\ave{\Cabs}$, $\ave{\Csca}$ and $\ave{\musca}$ of VO$_2$(M) bars of variable length and fixed width.} The bars dimensions are: $W = 2.5$ \microns~and $L =$ 5, 10, 15, 30 and 50 \microns. The refractive index of the host, $n_\mathrm{host} = 1.5$, and the refractive index of the VO$_2$(M) bars was obtained from the literature (see "film 2" in Wan et al\autocite{Wan2019}).}
\label{fig_S:scatter_vo2_bar_fixedW}
\end{figure}

The sensitivity of the VO$_2$(M) bars average light scattering to changes in $L$ for $W = 2.5$ \microns is shown in Fig.~\ref{fig_S:scatter_vo2_bar_fixedW}. As shown in the figure, $\ave{\Cabs}$, $\ave{\Csca}$ and $\ave{\musca}$ show small variation at $L > 15$ ~\microns. We noted a similar response for other values of $W$ (not shown here).

\subsection{Estimation of average light scattering of VO$_2$(M) microcrystal ensemble}\label{sec_S:ave_scatter_VO2_ensemble}

The average light scattering of the VO$_2$(M) microcrystal ensemble is calculated through:

\begin{subequations}\label{eq_S:ensemble_scatter}
\begin{align}
    \ave{\Cabs} &= \sum_{W,L} F_{W,L}\ave{\Cabs}_{W,L} + F_\mathrm{flake}\ave{\Cabs}_\mathrm{flake} \label{eq_S:ensemble_Cabs}
    \\
    \ave{\Csca} &= \sum_{W,L} F_{W,L}\ave{\Csca}_{W,L}  + F_\mathrm{flake}\ave{\Csca}_\mathrm{flake} \label{eq_S:ensemble_Csca}
    \\
    \ave{\musca} &= \frac{1}{\ave{\Csca}}\Bigg[\sum_{W,L}  F_{W,L}\ave{\Csca}_{W,L}\ave{\musca}_{W,L}
    + F_\mathrm{flake}\ave{\Csca}_\mathrm{flake}\ave{\musca}_\mathrm{flake}\Bigg]
    \label{eq_S:ensemble_musca}
\end{align}
\end{subequations}
where $F$ represent the number microcrystals of a particular morphology in the ensemble, and the subscripts "${W,L}$" and "$\mathrm{flake}$", indicate a microbars of dimensions $L$ and $W$, and flakes, respectively. For $F_{W,L}$, we considered the estimated size distribution [Main text, Fig.~\ref{fig:PE_VO2_composite}(c)], while  $F_\mathrm{flake} = 5$, as an approximate of the number of flakes found in the sample.

The volume of the microcrystal ensemble $V$ is given by:
\begin{equation}\label{eq_S:Vol_ensemble}
    V = \sum_{W,L}F_{W,L}V_{W,L} + 5V_\mathrm{flake},
\end{equation} 
\noindent where $V_{W,L}$ is the volume of a single bar with dimensions $W$ and $L$, and $V_\mathrm{flake}$ is the volume of the flake structure.

\subsection{Prediction of refractive index and scattering from PE films}\label{sec_S:PE_film_nk}

\begin{figure}[H]\centering
\includegraphics[width=1.0\textwidth]{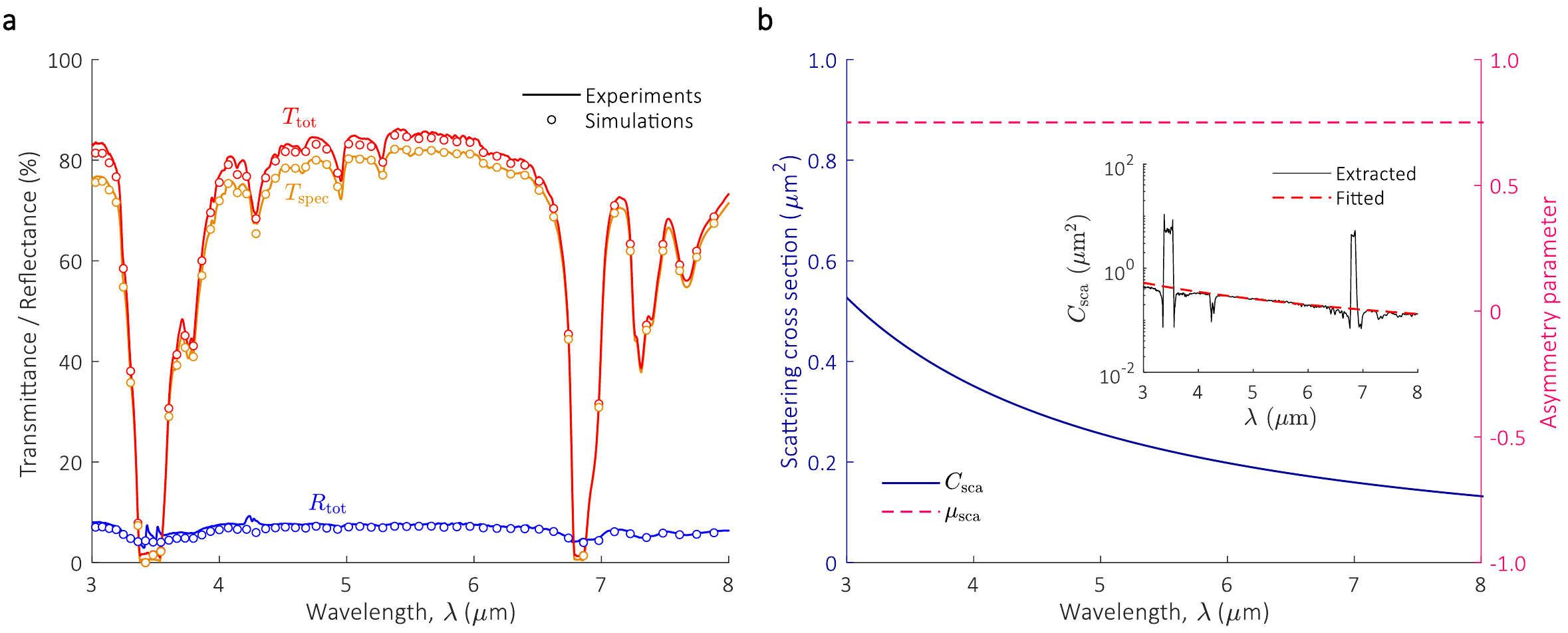}
\caption{\textbf{Estimation of scattering properties and refractive index of the PE films used in the VO$_2$(M)/PE composites.} \textbf{(a)} The values of $T_\mathrm{tot}$, $T_\mathrm{spec}$ and $R_\mathrm{tot}$ of a 101\microns~thick PE film obtained from experiments (solid lines) are compared with Monte-Carlo simulations based on the estimated refractive index and scattering properties of the PE film). \textbf{(b)} Estimated $\Csca$ of the PE film for $f_v = 0.1$\% v/v and particles of 1~\microns~diameter. The curve is based on a fitting curve using the extracted results from experiments (inset).}
\label{fig_S:PE_scattering}
\end{figure}

Overall, the optical properties of PE are highly dependent on the crystallinity and the manufacturing process\autocite{Palik1998}. Thus, we extracted the complex refractive index of a pure PE film (101~\microns~thick), $n_\mathrm{PE}$, fabricated under the same conditions than those used for the composite films. First [Fig.~\ref{fig_S:PE_scattering}(a)], $T_\mathrm{tot}$, $T_\mathrm{spec}$ and $R_\mathrm{tot}$ of the PE film where measured by FTIR and a gold integrating sphere (Main text, Materials and Methods). The real part of $n_\mathrm{PE}$ was obtained from Palik et al\autocite{Palik1998}. The extinction coefficient, $\kappa_\mathrm{PE} = \imag{n_\mathrm{PE}}$, was extracted from the equation\autocite{Katsidis2002a}:

\begin{equation*}
    T_\mathrm{tot} = \frac{\left(1 - R_f\right)^2\exp{\left(-2k_0t_\mathrm{film}\kappa_\mathrm{PE}\right)}}
                      {1 - R_f^2\exp{\left(-4k_0t_\mathrm{film}\kappa_\mathrm{PE}\right)}},
\end{equation*}
\noindent where $t_\mathrm{film}$ is the thickness of the PE film and $R_f$ is the reflectance at the air/PE interface. $R_f$ was estimated from Fresnel law and the refractive index of PE from Palik et al\autocite{Palik1998}.

As shown in Fig.~\ref{fig_S:PE_scattering}(a), $T_\mathrm{tot} \neq T_\mathrm{spec}$ at most parts of the spectrum, indicating a small scattering component in the film. To include this features in the modeling, we assume a small concentration of scattering particles inside the film. 
Because $\kappa_\mathrm{PE} \ll \real{n_\mathrm{PE}}$, $T_\mathrm{tot}\approx \left(1 - R_f\right)^2\exp{\left(-2k_0t_\mathrm{film}\kappa_\mathrm{PE}\right)}$\autocite{Palik1998}. Additionally, at low particle concentrations\autocite{Prahl1993}, $$T_\mathrm{spec}\approx \left(1 - R_f\right)^2\exp{\left[\left(-f_v\Csca/V_p-2k_0\kappa_\mathrm{PE}\right)t_\mathrm{film}\right]}.$$ Thus, the scattering properties of the particles are extracted from:

\begin{equation*}
    \Csca = -\frac{V_p\ln\left(T_\mathrm{spec}/T_\mathrm{tot}\right)}{f_vt_\mathrm{film}}.
\end{equation*}

The extracted values of $\Csca$ for $f_v = 0.1$\% v/v and particles of 1~\microns~diameter are shown at the inset of Fig.~\ref{fig_S:PE_scattering}(b). We fitted the results with a curve of the form $\Csca = a/\lambda^b$, with $a = 2.4782$ and $b = 1.4095$ [Fig.~\ref{fig_S:PE_scattering}(b)]. Finally, using Monte-Carlo simulations (see Main Text, Materials and Methods) we iterate to find the value of $\musca$ that best matches the results of experiments. We found excellent agreement for $\musca = 0.75$ [Fig.~\ref{fig_S:PE_scattering}(a)]. 

\subsection{Radiative properties of VO$_2$(M)/PE composite films (Experiments and Simulations}\label{sec_S:PE_VO2_composite_extra}

\begin{figure}[H]\centering
\includegraphics[width=1.0\textwidth]{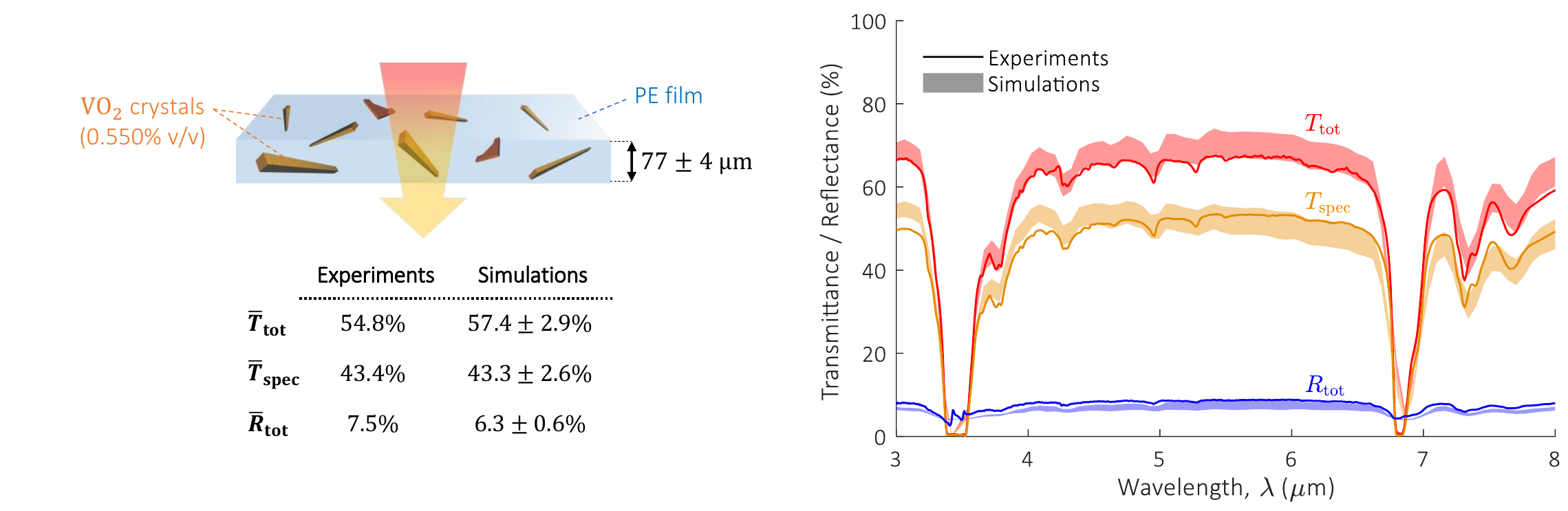}
\caption{\textbf{$T_\mathrm{tot}$, $T_\mathrm{spec}$ and $R_\mathrm{tot}$ of VO$_2$(M)/PE composite film obtained from simulation and experiments}. The composite film is composed of a PE matrix with 0.55\% v/v VO$_2$(M) powder and has $77\pm4$\microns~thickness. The scattering properties of the VO$_2$(M) powder are the same used in the main text [Fig.~\ref{fig:PE_VO2_validation}(a)]}
\label{fig_S:PE_VO2_composite_extra}
\end{figure}

\section{Validation of Monte-Carlo Code}\label{sec_S:Monte-Carlo}

\begin{figure}[H]\centering
\includegraphics[width=1.0\textwidth]{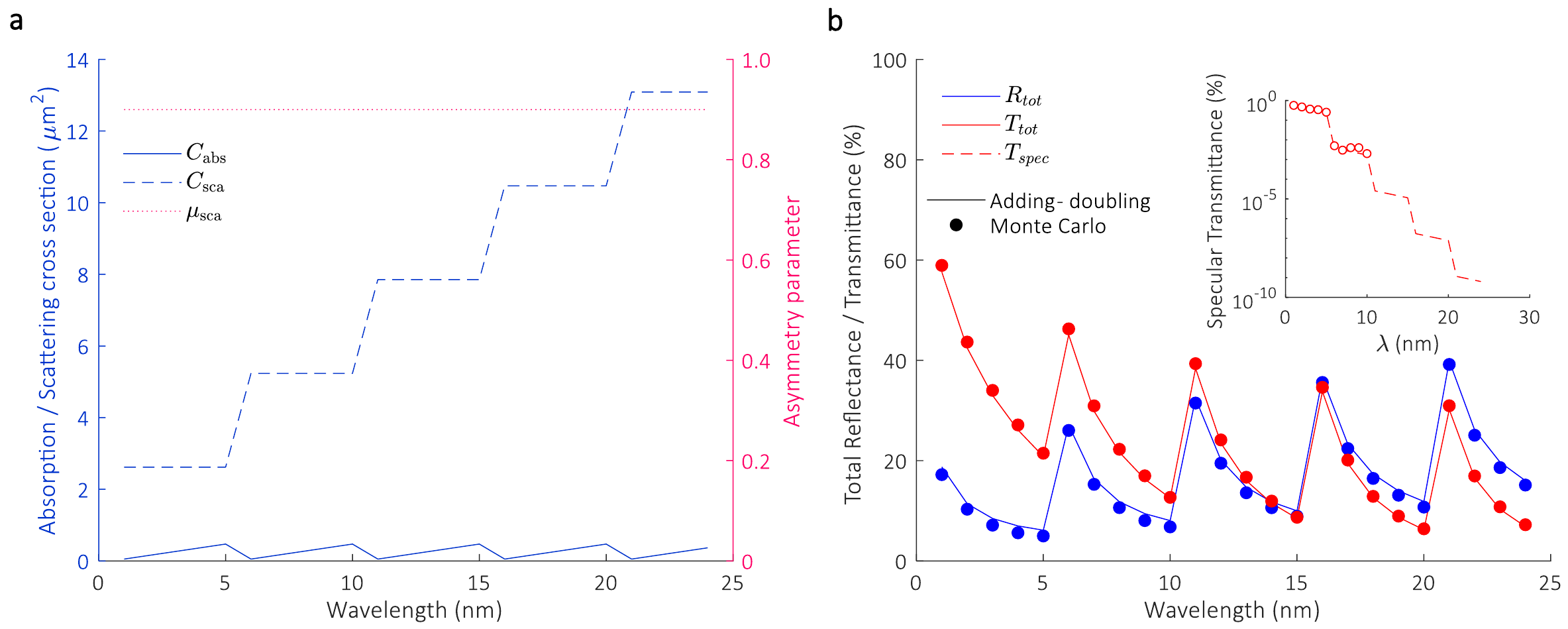}
\caption{\textbf{Validation of Monte-Carlo code for radiative transfer calculations against Adding-doubling method}. \textbf{(a)} Light scattering properties $\Cabs$, $\Csca$ and $\musca$ of the spherical particles (1 \microns~diameter) considered in this study \textbf{(b)} Total transmittance and reflectance and specular transmittance (inset) for light at normal incidence of a 1 mm thick film with 1.4 refractive index and 0.1\% particles per volume. The results from our Monte-Carlo code (circles) are plotted against Adding-doubling method calculations using the code by Prahl\autocite{iad_code2021}.}
\label{fig_S:monte_carlo_validation}
\end{figure}

We validated our Monte-Carlo code for radiative transfer simulations against the Adding-doubling method [Fig.~\ref{fig_S:monte_carlo_validation}(b)]. We calculated the total transmittance ($T_\mathrm{tot}$) and reflectance ($R_\mathrm{tot}$), and the specular transmittance ($T_\mathrm{spec}$) at normal incidence over a 1 mm thick film with refractive index $n_\mathrm{film} = 1.4$, and 0.1\% particles per volume. The particles have a diameter $D_p = 1$ \microns, and the scattering properties shown in Fig.~\ref{fig_S:monte_carlo_validation}(a).

The results for Adding-doubling method are obtained from the open-source code by Prahl\autocite{iad_code2021}. This code was used to calculate the total reflectance and transmittance. The specular transmittance is calculated as $$T_{\mathrm{spec}} = (1 - R)\exp\left[\frac{f_v}{V_p}\left(\Cabs + \Csca\right)t_\mathrm{film}\right],$$ where $R = 4.12\%$ is the reflectance of a 1 mm thick film with 1.4 refractive index, $f_v$ is the volume fraction, $V_p$ is the particle's volume, and $t_\mathrm{film}$ is the thickness of the film.

The results of total transmittance/reflectance from our Monte-Carlo code show excellent agreement with the curves from Adding-doubling method. The specular transmittance (inset of Fig.~\ref{fig_S:monte_carlo_validation}) shows good agreement up to 10 nm wavelength. Outside this range, the values of specular transmittance are bellow the minimum resolution of our Monte-Carlo setup (0.0001\% for 1,000,000 photons).

We performed and additional test based on the total transmittance and reflectance of a slab with particles, considering the conditions: $n_\mathrm{film} = 1.0$, $t_\mathrm{film} = 200$ \microns, $D_p = 1$ \microns, $f_v$ = 0.1\%, $\Cabs = 0.5236$ $\mu$m$^2$, $\Csca = 4.7124$ $\mu$m$^2$ and $\musca = 0.75$. The results shown in Table~\ref{table_S:montecarlo_test1}, show excellent agreement against the exact solution from Van de Hulst\autocite{VandeHulst1980} and simulations with two different Monte-Carlo codes\autocite{Jacques1995,Prahl1989}.

\begin{table}[H]
\centering
\begin{tabular}{ |c|c|c|c|c| }
\hline
 $R_\mathrm{tot}$ & $T_\mathrm{tot}$ & Source \\
 \hline
 9.739\% & 66.096\% & Van de Hulst\autocite{VandeHulst1980} \\
 9.774\% & 66.101\% & Monte Carlo code \\
 9.734\% & 66.096\% & Wang, Jacques and Zheng\autocite{Jacques1995} \\
 9.711\% & 66.159\% & Prahl et al\autocite{Prahl1989} \\
 \hline
\end{tabular}
\caption{\textbf{Validation of Monte-Carlo code for Test 1.} Total transmittance and reflectance of a slab with particles, considering the conditions: $n_\mathrm{film} = 1.0$, $t_\mathrm{film} = 200$ \microns, $D_p = 1$ \microns, $f_v$ = 0.1\%, $\Cabs = 0.5236$ $\mu$m$^2$, $\Csca = 4.7124$ $\mu$m$^2$ and $\musca = 0.75$}
\label{table_S:montecarlo_test1}
\end{table}

\printbibliography

\makeatletter\@input{xx_Manuscript.tex}\makeatother